\documentclass[prb,twocolumn,superscriptaddress,amsmath,amssymb]{revtex4}

\newcommand{\fatr}{\mathbf{r}}

\usepackage{array}
\usepackage{graphicx}
\usepackage{color}
\usepackage{bm}

\def\beq{\begin{equation}}
\def\eeq{\end{equation}}
\def\bea{\begin{eqnarray}}
\def\eea{\end{eqnarray}}
\def\fatR{{\bf R}}
\def\fatr{{\bf r}}
\def\fatM{{\bf M}}

\begin{document}
\title{First principles view on chemical compound space: Gaining rigorous atomistic control of molecular properties} 
\author{O.~Anatole von Lilienfeld}
\email{anatole@alcf.anl.gov}
\affiliation{Argonne Leadership Computing Facility, Argonne National Laboratory, Argonne, Illinois 60439, USA}

\date{\today}

\begin{abstract}
A well-defined notion of chemical compound space (CCS) is essential for gaining rigorous control of properties 
through variation of elemental composition and atomic configurations. 
Here, we review an atomistic first principles perspective on CCS. 
First, CCS is discussed in terms of variational nuclear charges in the context of
conceptual density functional and molecular grand-canonical ensemble theory.
Thereafter, we revisit the notion of compound pairs, 
related to each other via ``alchemical'' interpolations involving fractional nuclear chargens in the electronic Hamiltonian. 
We address Taylor expansions in CCS, property non-linearity, improved predictions using 
reference compound pairs, and the ounce-of-gold prize challenge to linearize CCS.
Finally, we turn to machine learning of analytical structure property relationships in CCS.
These relationships correspond to inferred, rather than derived through variational principle, 
solutions of the electronic Schr\"odinger equation.
\end{abstract}

\maketitle

\section{Introduction}
In analogy to the vastness and sparseness of outer space, we can loosely refer to the space of chemical
systems as chemical compound space (CCS), i.e.~some continuous observable space that is populated 
by all epxerimentally {\em and} theoretically possible chemicals with integer nuclear charges
and interatomic distances for which chemical interactions occur~\cite{ChemicalSpace}.
Stated more precisely, CCS refers to the combinatorial set of all compounds that can be isolated and constructed 
from possible combinations and configurations of $N_I$ atoms and $N_e$ electrons in real space.
In absence of external fields and given $N_e$ and $N_I$ atom-types $\{Z_I\}$ and spatial configurations $\{\fatR_I\}$,
not only covalent, ionic, and metallic bonding result, but 
also the much weaker hydrogen and van-der-Waals-bonding, responsible for the physics and chemistry of
molecular crystals, liquids, and other supra-molecular aggregates, can be derived,
as well as all other quantum and statistical mechanical properties such as electronic states, electronic and vibrational spectra, 
free energies, and even phase diagrams and rare events such as chemical reactions.
While most research efforts in this first principles context have been dedicated to 
the approximations and methods necessary for making property predictions for given compounds,
the focus of this tutorial is a first principles view on the compounds {\em per se}. 

Notwithstanding chemical bonding or conformations and merely considering the number of possible stoichiometries 
it is obvious that the size of CCS is unfathomably large for all but the smallest systems. 
Due to all the possible combinations of assembling many and various atoms
its size scales exponentially with compound size as $\propto Z_{max}^{N_I}$. 
Here $Z_{max}$ is the number of possible atom types, i.e. the maximal permissible nuclear charge in Mendeleev's table ($Z_{max} > 100$), 
and $N_I$ depends on the employed definition of ``isolated system'' but can certainly reach 
Avogardro's number scale for living organisms, chunks of unordered matter, or planets.
While many of such speculative compounds are likely to be unstable, the state of affairs 
worsens when accounting for the additional degrees of freedom which arise from
distinguishable geometries due to differences in atom bonding or conformations. 
This combinatorial explosion with system size is the main motivation for advocating an {\em ab initio}, or first principles, 
view on CCS, i.e.~a view that restricts us to use solely $\{Z_I\}$ and $\{\fatR_I\}$ as input 
variables~\footnote{Include the mass of the nuclei as an additional variable if also dynamical properties and nuclear quantum effects are to be accounted for.}, and, while maybe not free of parameters, will not change in its parameterization as $\{Z_I\}$ and $\{\fatR_I\}$ are freely
varied~\cite{AbInitioDefinitionByKieronBurke}.
A major part of modern electronic structure theory and interatomic potential work is concerned with 
the development of improved methods and approximations for solving Schr\"odinger's equation (SE) within the
Born-Oppenheimer approximation for Hamiltonians relevant to materials, biological, or chemical
research, and deriving properties thereof~\cite{MolecularElectronicStructureTheory}.
{\em Ab initio} statistical mechanics efforts are dedicated to sampling the corresponding 
$3N_I-6$ degrees of freedom from first principles~\cite{tuckerman_book_SM}.
In the context of CCS, the electronic Hamiltonian $H$ for solving SE, 
$H\Psi = E\Psi$, of {\em any} compound with a given charge, $Q = N_p - N_e$,
is uniquely determined by its (unperturbed) external potential, $v(\fatr) = \sum_I Z_I/|\fatr-\fatR_I|$, i.e.~by its set $\{\fatR_I,Z_I\}$.
Here, $N_p$ is the total number of protons in the system, i.e.~the sum over all nuclear charges.
Due to the Hohenberg-Kohn theorem we also know that the electron density $n(\fatr)$, and
all electronic properties derived thereof, are determined by $\{Z_I,\fatR_I\}$, up to a trivial constant,
$\{H(\fatr),N_e\} \leftrightarrow \{Z_I,\fatR_I,N_e\} \leftrightarrow \{v(\fatr),N_e\} \leftrightarrow n(\fatr)$~\cite{HK}.
Consequently, we work with $\{Z_I,\fatR_I,N_e\}$.

In this tutorial, CCS is first briefly illustrated in terms of a rough energy scale in section~\ref{sec:CCS}. 
In section~\ref{sec:MGCE} we will review the notion of a molecular grand-canonical ensemble density functional theory that 
can deal with fractional electrons {\em and} nuclear charges. 
Section~\ref{sec:PAIRS} will deal with pairs of chemical compounds, and with efforts
to exploit the arbitrariness of interpolating functions. 
It also details the challenge associated to a prize award of one ounce of gold.
Finally, we will study recent efforts to use intelligent data analysis methods (machine learning)
to systematically infer analytical structure property relationships from previously calculated 
electronic structure data sets in section~\ref{sec:ML}.

\begin{figure}[ht]
\centering
\includegraphics[scale=0.4, angle=0]{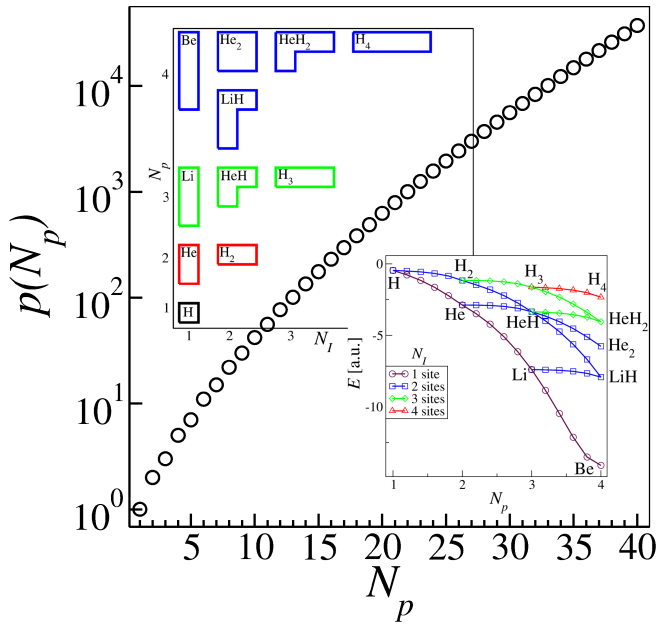}
\caption{
Exponential scaling of the total number of all possible $N$ partitions, i.e.~stoichiometries,
as a function of $N_p$ (Regime (i) in section~\ref{sec:CCS}).
Inset upper left-hand side: Young-Ferrers diagrams illustrating the possible 
partitions (stoichiometries) for $N_p$ as a function of number of atoms, $N_I$.
The color code corresponds to the total number of protons in the compound, $N_p \in  \{$1 (black), 2 (red), 3 (green), 4 (blue)$\}$ 
Inset lower right-hand side: Total potential energy of relaxed molecules as a function of $N_p$ using
interpolated pseudopotentials in analogy to Fig.~\ref{fig:PPs} 
(all systems neutral, BLYP DFT level of theory~\cite{B88X,lyp}, arbitrary energy origin due to use of pseudopotentials). 
}
\label{fig:partition}
\end{figure}

\section{Energy hierarchy}
\label{sec:CCS}
Regarding CCS it is useful to think of a variable system that is comparable to the Mendeleev's table of the elements.
Compounds, however, have many more dimensions than a single atom's nuclear charge, specifically $4N_I - 6$ ($4N_I -5$ if linear).
One way of thinking about CCS is in terms of an abstract {\em Gedankenexperiment} involving all theoretically existing compounds.  
Consider all the compounds possible for a set of protons, subject to a varying amount
of kinetic energy (or temperature), provided by a thermostat. 
Regimes emerge of various familiar degrees of freedom.
In this ``phase diagram'' of CCS these various regimes correspond to 
\begin{itemize}
\item[(i)] {\em stoichiometrical isomers}: A fictitious ``very high'' temperature regime. 
Let us assume such high temperatures that all bonds break, and that all spatial degrees of freedom can safely be neglected.
Furthermore, we assume isomers to have the same number of elementary particles, $N_p$ and $N_e$.
How many of such stoichiometrical isomers could be observed populating up to $N_I \le N_p$ sites with at least one proton?
Mathematically speaking, this is a discrete number theory problem: 
This number is the integer partition of $N_p$, i.e.~the number of ways to write $N_p$ as sum of positive integers.
For example, CH$_4$, NH$_3$, H$_2$O, HF, Ne represent only 5 out of all the 42 possible stoichiometrical isomers for $N_p = N_e =$ 10.
The total number of possible partitions corresponds to the partition function, which increases exponentially with $N_p$. 
The exponential increase, and an illustration of the emerging stoichiometries according to Young-Ferrers diagrams,
are shown in Fig.~\ref{fig:partition}.
These degrees of freedom are rarely explored in nature except when it comes to radioactive decay, nuclear fusion, 
or nuclear synthesis in the early stages of our universe.
Through interpolation, however, we can meaningfully render this space continuous, as illustrated using 
density functional theory (DFT) for the potential energies displayed in the inset of Fig.~\ref{fig:partition}. 
\item[(ii)] {\em constitutional isomers}: ``high'' temperature regime. 
At high temperatures, only strong chemical bonding (covalent, ionic, metallic) survives.
Corresponding Lewis structures enumerate many (but not all) of the possible constitutional 
isomers distinguishable as possible topologies, or molecular graphs, that can be constructed.
The enumeration (and canonization) of all possible constitutional isomers has been the focus of long standing
graph-theoretical efforts~\cite{EnumerateAlkene,EnumerateAlkene2,EnumerateAlkene3,CanonizerMeringer2004}.
The exponential scaling of their number is also evident for the recently published exhaustive
list of small organic molecules~\cite{ReymondChemicalUniverse3}.
This is the regime in which isomerism occurs through conventional ``chemistry'', 
i.e.~reactions that lead from one constitutional isomer to another, 
usually under the influence of pressure, temperature, light, or in the presence of some catalytic agent.
We can model this and the subsequent regimes using {\em ab initio} molecular dynamics methods~\cite{tuckerman_book_SM}.
Universal or reactive force-fields attempt to accomplish similar sampling~\cite{ReaxFF2001,UFFRappeGoddard1992}.
\item[(iii)] {\em conformational isomers}: ``ambient'' temperature.
Folding and un-folding events, sampling of intramolecular degrees of freedom, for example
around dihedral angles and similar processes take place at ``ambient'' temperatures.
These isomers are typically sampled using force-fields that assume fixed molecular topologies and
parameterized charges, dihedral and angular terms, in addition
to the typical potentials used for the chemical bond, such as harmonic, Buckingham's or Morse potentials.
\item[(iv)] {\em weakly interacting systems}: ``low'' temperature.
Supra-molecular assemblies, soft aggregates condense to molecular liquids or solids. 
Typically modeled using classical effective Lennard-Jones type potentials.
\end{itemize}

In the remainder of this review we will discuss recent contributions that are consistent with {\em all}
of the four regimes, accounting for all the spatial {\em and} elemental degrees of freedom $\{\fatR_I\}$ and $\{Z_I\}$.
As also discussed below, we will ignore the electronic number as an independent variable 
since $N_e \approx N_p = \sum_I Z_I$ for most if not all possible scenarios.

\section{Molecular grand-canonical ensemble}
\label{sec:MGCE}
\subsection{Theory}
Much of conceptual density functional theory (DFT) concerns the energy response to infinitesimal variations
in number of electrons $N_e$, or external potential, $v(\fatr)$~\cite{Geerlings_DFTConcepts,parryang}. 
While very important for interpreting orbitals, deriving reactivity indices, and even for redox-processes, 
the diversity (and combinatorial scaling) of CCS is rather due to variations 
in nuclear charge distribution, than due to variations in $N_e$.
Consequently, for the following we will mostly be concerned with changes in nuclear charges. 
In order to offer a rigorous framework for explicit changes in $\{Z_I\}$, 
molecular grand-canonical ensemble DFT was introduced~\cite{anatole-jcp2006-2}, 
relying to a significant degree on preceding work~\cite{nonBO_DFT,anatole-prl2005}.
Only a brief summary is given here, for more details the reader is referred to the original contributions.

Assuming a classical nuclear charge distribution, $n_p(\fatr)$, 
we can introduce an auxiliary grand-canonical variational energy functional for the aforementioned
fictitious ``very high'' temperature regime (i),
\bea
\Omega[N_e,n_p(\fatr)] & = & E[N_e,n_p(\fatr)] - \mu_e \left(\int d\fatr\; n(\fatr) - N_e\right) \nonumber\\
& & - \mu_p \left( \int d\fatr \; n_p(\fatr) - N_p\right). 
\eea
Where $E,n,\mu_e,\mu_p$ correspond to the usual total potential energy functional, the electron charge density, 
and global electronic and nuclear chemical potentials, respectively.
For high temperatures, entropy will prevail and the system would dissociate into H$_{N_p}$. 
For lower temperatures, the potential energy will dominate the free energy, 
and the energy of a single atom ($E(Z) \approx Z^{2.4} = N_p^{2.4}$) will dominate 
over the energy of many individual atoms that sum up to the same number of protons, $\sum_I Z_I^{2.4} \le N_p^{2.4}$.
Hence, the nuclear charge distribution would collapse onto a single site. 
For this discussion, we assume that the classical and fictitious self-repulsion of protons occupying 
the same nuclear site is switched off. 

For the lower temperature regimes (ii)-(iv), the following energy functional is more meaningful,
\bea
\label{eq:Omega}
\Omega[N_e,n_p(\fatr)] & = & E[N_e,n_p(\fatr)] - \mu_e \left(\int d\fatr\;\; n(\fatr) - N_e\right) \\
& & - \int d\fatr \; \mu_p(\fatr) \left(n_p(\fatr) - \sum_I Z_I \delta(|\fatr-\fatR_I|)\right) \nonumber
\eea
where $\sum_I Z_I \delta(|\fatR_I -\fatr|)$ corresponds to the spatially resolved nuclear charge distribution. 
The nuclear chemical potential $\mu_p$ is no longer a global parameter but rather a locally defined Lagrange multiplier. 
Using an external potential that excludes the aforementioned intra-nuclear self-repulsion of protons (here through use of an error function),
we find for the corresponding Euler equation, 
\bea
\mu_p(\fatr) & = & \frac{\delta E}{\delta n_p(\fatr)}  
\;\;=\;\; \sum_I \frac{Z_I {\rm erf}[\sigma|\fatR_I-\fatr|]}{|\fatr-\fatR_I|} - \int d\fatr' \; \frac{n(\fatr')}{|\fatr-\fatr'|},
\nonumber\\
\eea
---the electrostatic potential of the system.
As such, starting with $Q$---a Legendre transformed energy functional of intensive properties $\mu_e$ and $\mu_p(\fatr)$---one 
can derive the Gibbs-Duhem equation for electrons {\em and} protons and electrons,
\bea
dQ[\mu_e,\mu_p(\fatr)] & = & -N_e \; d\mu_e - \int d\fatr\; n_p(\fatr)\; \delta \mu_n(\fatr), 
\eea
and obtain relationships between electronic hardness $\partial^2 E/\partial N_e$, 
molecular Fukui function $\partial \mu_p(\fatr)/\partial N_e$~\cite{anatole-jcp2007}, 
and nuclear hardness, $\delta \mu_p(\fatr)/\delta n_p(\fatr)$~\cite{anatole-jcp2006-2}.
While the nuclear chemical potential is defined everywhere, its value at an atomic position quantifies the system's first
order energy response to a fractional change of the atom's nuclear charge. 
Consequently, we dub $\mu_p(\fatR_I)$ the ``alchemical potential'' of atom $I$~\cite{anatole-prl2005}.

\subsection{Interpolating pseudopotentials}

\begin{figure}[ht]
\centering
\includegraphics[scale=0.4, angle=0]{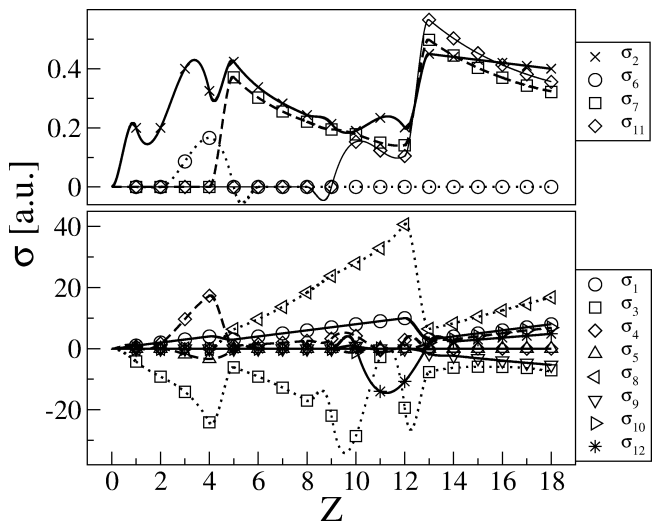}
\caption{
Interpolation of Goedecker-Hutter pseudopotential parameters for the BLYP based DFT calculations~\cite{SG,KrackPP}.
Parameters, shown as a function of nuclear charge, become polynomial regressions of third degree in $Z$ for intervals of 
$Z$ connected such that the sum is continuous and differentiable everywhere. 
}
\label{fig:PPs}
\end{figure}

Apart from radioactive processes alchemical changes obviously do not occur in reality. 
They offer, however, a rigorous mathematical way to render CCS continuous. 
Alchemical changes and potentials involving fractional nuclear charges are commonly 
used for two, often related, purposes: Either for the evaluation of free energy differences
between different compounds, e.g.~using thermodynamic integration~\cite{TI}, $\Delta F = \int d\lambda \; \langle \partial E/\partial \lambda \rangle$;
or for obtaining a set of gradients with dimension of $N_I$ indicating 
the response of the system to a variation in nuclear charge on every site~\cite{anatole-prl2005,anatole-jctc2007}.
In practice, we can calculate such changes through interpolation of nuclear charges in any basis set that is converged
for all values of an interpolating order parameter, $\lambda$.
For plane-wave pseudopotential implementations, the same can be accomplished by interpolation of pseudopotentials 
that replace the explicit treatment of the core electrons~\cite{Hellmann1,Hellmann2,KleinmansPP,WeeksPP,bhs-with-title,ChristiansensPP}.
The use of a plane-wave basis set is advantageous since it is independent of atomic position and type,
and will not introduce Pulay forces~\cite{Pulay-force}.
The manipulation of pseudopotentials for affecting electronic structure properties is nothing out of the ordinary.
It has successfully been deployed for an array of properties including relativistic effects~\cite{sgpsp}, 
self-interaction corrections,~\cite{SICPP-Vogl,SICPP-Pollmann}
exact-exchange and QM/MM boundary effects~\cite{anatole-jcp2005,Christiansen-JCP-2002},
van-der-Waals interactions~\cite{anatole-prl2004,DiLabioDCACP2012}, and
widening the band gap~\cite{Christensen1984,vandewalle2007}.
For fractional nuclear charges we can interpolate pseudopotentials, and evaluate properties as a function 
of order parameter, $0 \le \lambda \le 1$.
An interpolation of pseudopotential parameters as a function of nuclear charge is shown in Fig.~\ref{fig:PPs}. 
Calculated properties as a function of such alchemical changes are illustrated in 
Figs.~\ref{fig:partition} and \ref{fig:HF} for total potential energies, and protonation energies and polarizabilities, respectively.
Note that the former property is not physical because of the arbitrary energy offset of pseudopotentials. 
This, however, is inconsequential, since most of chemistry deals with energy differences, and differences thereof. 
As shown in Fig.~\ref{fig:HF} for HCl $\rightarrow$ NH$_3$, 
the use of pseudopotentials for alchemical changes can be particularly advantageous when it comes to transmuting
elements from different rows of the periodic table while keeping constant the total number of {\em valence} electrons. 

\subsection{Free energy applications}
Fractional charges were used to calculate free energy of solvation of ions in water~\cite{anatole-jcp2009}.
Sulpizi and Sprik rigorously explored the need for fractional nuclear charge calculations to obtain pK$_a$'s of 
various organic and inorganic acids and bases~\cite{pKaLore2008}.
In the case of free energy differences, fractional charges can also be avoided all together within a simple alternative
and elegant interpolation scheme put forth by Alf\`e, Gillan and Price:
Atomic forces are evaluated at both end-points ($\lambda \in {0,1}$), 
and $\lambda$ dependent molecular dynamics trajectories are generated for atoms being propagated 
according to a linear combination of these forces using instantaneous 
$\lambda$ values as weights~\cite{Alfe-nature2000}. 
This is to be compared to a trajectory that uses
Hellmann-Feynman forces directly evaluated on the interpolated alchemical species, 
such as for the 0 Temperature limit of relaxing the geometry of a reaction barrier~\cite{CatalystSheppard2010}. 
The limitations of Alf\`e's procedure are that (a) one requires twice as many self-consistent field calculations, 
namely for both end-points instead of a single one when using an alchemical interpolation
(assuming of course that for both approaches the free energy integrand $\langle \partial E/\partial \lambda \rangle$, 
varies similarly with $\lambda$); and (b) that the number of atoms must be kept constant during the interpolation, 
significantly restricting the possible number of stoichiometries that can be explored. 
Both of these disadvantages can be avoided within the compound pair scheme discussed in section~\ref{sec:PAIRS}.

\subsection{Design applications}
Through use of a Taylor expansion, truncated after first order, we can also
exploit the vector of alchemical potentials for gaining control over properties in 
the $N_I$-dimensional space of all the nuclei that are adjacent in the periodic table. 
Weigend, Schrodt and Ahlrichs were probably the first to present such an application for
the prediction of stability in binary atom clusters~\cite{AlchemicalDerivativeBinaryMetalCluster_WeigendSchrodtAhlrichs2004}.
Subsequently it was applied to drug binding energies within a QM/MM study~\cite{anatole-prl2005},
demonstrated for hydrogen-bonded
complexes, inter-converting methane, ammonia, water and hydrogen-fluoride while bound to formic acid~\cite{anatole-jctc2007},
and applied to the molecular Fukui function for tuning HOMO eigenvalues of boron/nitride derivatives of benzene~\cite{anatole-jcp2007}.
In Ref.~\cite{CatalystSheppard2010} this notion is exploited for the prediction of reaction barriers
as well as oxygen adsorption energies on Pd$_{79}$ derived core-shell metal nano-clusters that
are catalyst candidates for the oxygen reduction reaction.
The molecular Fukui function, in particular when evaluated at the position of the atom, was 
also discussed more recently by Cardenas et al.~\cite{CardenasFukui2011}.
Clearly, truncation of the Taylor expansion at second or higher order would be desirable to
increase the accuracy of the alchemical predictions of the effect of atomic transmutations.
Alas, higher order derivatives of the energy with respect to nuclear charges lead to computational overhead, 
they require the calculation of the perturbed electronic structure. 
Nevertheless, based on coupled perturbed self-consistent field theory
the improved accuracy of higher order predictions was demonstrated very recently~\cite{LesiukHigherOrderAlchemy2012}.

\subsection{Other applications}
Instead of varying the proton distribution $n_p(\fatr) = \sum_I Z_I \delta(|\fatr-\fatR_I|)$ in a compound, 
we can interpolate the external potential $v(\fatr) = \sum_I Z_I/|\fatr-\fatR_I|$ just as well.
Albeit mixing up spatial and compositional degrees of freedom, this is more in line with conventional thought
in quantum chemistry and conceptual DFT~\cite{Geerlings_DFTConcepts} where the nuclear charge distribution is hardly mentioned explicitly.
The route via the external potential has been pursued within the {\em Ansatz} of linear 
combination of atomic potentials in the research groups of 
Yang and Beratan~\cite{RCD_Yang2006}, assigning atom-type specific weights to every atom site. 
Using simplified Hamiltonians, impressive results were obtained for the control of
molecular hyper-polarizabilities~\cite{InverseBeratanYang2008,GradientMCBeratanYang2008,HybridExplorationCCSBeratanYang2008,DesignKeinanBeratanYang2008,MolecularDesignRinderspacherBeratanYang2009}, based on long-standing molecular design
efforts for electronic properties well ahead of their time~\cite{Beratan1991,Beratan1996}.
This approach has also been adapted and explored for the purpose of crystal structure design using DFT~\cite{AvezacZunger-prb2008}.
The functional second order derivatives with respect to external potentials have been published in 
Ref.~\cite{2ndOrderDerivsWrtVext_DeProftAyersGeerlings}.
Analytical expressions for second order derivatives and linear response functions have very recently been
proposed by Yang, Cohen, De Proft and Geerlings~\cite{YangCohenGeerlingsAnalyticalFukui2012}.
The same authors also derived important constraints for the electronic structure that must be met by the exact 
exchange-correlation functional. 
In analogy to using constraints obtained for variable $N_e$, such as piece-wise 
linear behavior and derivative discontinuities, in order to design improved
density functionals~\cite{DD_Perdew1,DD_Perdew2,DiscontinuousXC_MoriSanchezCohenYang2009}, 
Cohen's current efforts are dedicated to variations in the external potentials that include fractional nuclear charges.
The electronic structure for systems with $Z \rightarrow \infty$ has also been explored by Constantin et al.~\cite{HighZ_Burke2010}.

\section{Compound pairs}
\label{sec:PAIRS}
\subsection{Background}
The above discussed {\em Ansatz}, variational in a fractional nuclear charge distribution, 
defines an appealing, fully spatially resolved, index, i.e.~a way to probe the sensitivity of a compound not only 
towards changes in any of its composing atoms but also with respect to adding new protons. 
However, for two reasons this approach can also be limited. 
First, severe constraints and preconceived insights are required to explore the $N_I$-dimensional space of all $\{Z_I\}$. 
Either because if $Z_I$ is continuous it requires a bias potential towards integer numbers, possibly using a 
fictitious temperature, i.e.~in analogy to the Fermi function for electrons.
Or if $Z_I$ is a combination of various atom-types, i.e.~in line with the aforementioned
linear combination of atomic potentials approach~\cite{RCD_Yang2006}, 
the weight of one
nuclear charge has to dominate so that it can safely be increased to 1, while decreasing all others to zero.
Furthermore, constraints due to overall charge conservation, and electronic structure, have to be considered.
For example, consider an alchemical transmutation of H$_2$N-OH into its iso-electronic stoichiometrical isomer hydrogen
peroxide, HO-OH, through simultaneously and continuously decreasing and 
increasing by one the nuclear charge of the hydrogen and nitrogen atom, respectively.
At some point of this conversion, the spin of the ground state surface will turn into a triplet surface, therefore requiring the
consideration of {\em both} spin states along the interpolation path.
Second, and more importantly, in order to carry out alchemical changes along columns in the periodic table, 
for example, a path following $Z$ would have 
to fill up the shell to go through the entire period before one arrives at the desired elements.
This implies significant variations in electronic configurations just to arrive at a target compound with a 
configuration likely to be very similar to the starting compound. For example, consider a system of 8 valence electrons, and Ne and Ar as 
starting and target compounds, respectively. Then, an iso-electronic path progressing with $Z$ of the central atom, 
and saturating with hydrogens accordingly, would have to proceed through the following series of compounds, 
NaH$_7$, MgH$_6$, AlH$_5$, SiH$_4$, PH$_3$, H$_2$S, and HCl, some of which not even likely to be covalently bound.
Hence, while Taylor expansions in $Z$ are quite predictive for adjacent elements---as mentioned in the preceding section--- 
it is not surprising that their predictive power decays dramatically when it comes to predictions for changes up and down
the columns in the periodic table.
Obviously, matters only become worse when $d$- or $f$-elements have to be included, or when trying to 
make predictions by 2 or more rows down or upward. 

\begin{figure}[ht]
\centering
\includegraphics[scale=0.32, angle=0]{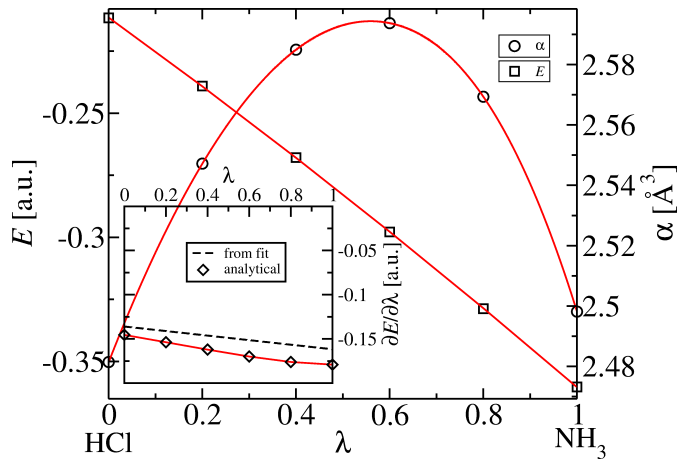}
\caption{
Protonation energy, $E = E^{\rm X-H^+} - E^{\rm X}$, and static polarizability $\alpha$ of neutral species, 
as a function of order parameter $\lambda$ driving X = HCl into X = NH$_3$. 
The inset shows the derivative of the protonation energy, once evaluated analytically according
to Hellman-Feynman in Eq.~(\ref{eq:HF}), and once from a quadratic fit to protonation energy.
Heavy atoms of two end-point molecules were superimposed, their effective nuclear charge being
($1-\lambda$)7+$\lambda$ 5, and hydrogens were placed in $xy$-plane. 
The protonating proton was placed in $z$, 1 {\AA} above the heavy atom.
All energies calculated with PBE functional and interpolated analytical pseudopotentials~\cite{sgpsp}.
}
\label{fig:HF}
\end{figure}

\subsection{Theory}
Albeit intuitive, the use of nuclear charges as interpolating variable is fortunately not mandatory.
Instead, we can also use a generalized, and entirely arbitrary, interpolation procedure between any two pairs
of compounds---as long as it is reversible 
and integrable, {\em any} path can be used to monitor any property that is a state function~\cite{tuckerman_book_SM}. 
In all of the following, we will only consider interpolations between iso-electronic compounds, i.e.~compounds with the same 
$N_e$ in their Hamiltonian.
As mentioned before, this is only a minor restriction because the diversity of CCS is by far not as much due to differences 
in $N_e$ as it is due to differences in nuclear charge distribution.
We can directly interpolate the nuclear charge distributions/external potentials/Hamiltonians of any two such
iso-electronic compounds, $A$ and $B$, e.g.~
\bea
H(\lambda,\fatr) & = & H_A(\fatr) + \lambda \left(H_B(\fatr) - H_A(\fatr)\right),
\label{eq:Hamiltonian}
\eea
interpolating linearly and globally in order parameter $\lambda$.
$H_A$ and $H_B$ denote the initial and final electronic Hamiltonian of the two compounds,
defining the corresponding boundary conditions $H(\lambda=0)$ and $H(\lambda=1)$, respectively. 
For any iso-electronic Hamiltonian linear in $\lambda$, the potential energy is not necessarily linear.
In fact, the electronic potential energy of a linearly interpolated Hamiltonian 
will always be concave, i.e.~equals or larger than a straight line connecting the energies of compound $A$ and $B$,
$E^{lin} = E_A-\lambda(E_B-E_A)$.
This inequality follows from the variational principle and can easily be shown: 
Eq.~\ref{eq:Hamiltonian} implies,
\bea
E[H(\lambda)] & = & \langle H(\lambda) \rangle_\lambda \;\;= \;\;  E_A[n_\lambda] + \lambda(E_B[n_\lambda] - E_A[n_\lambda]),\nonumber\\ 
\eea
where $\langle ... \rangle_\lambda$ now correspond to the usual quantum mechanical Bra-Ket notation, denoting
the expectation value with the wavefunction, or the density functional (in an orbital-free exact DFT world), 
evaluated for the Hamiltonian at $\lambda$, i.e.~$E[H(\lambda)] = \langle \Psi_\lambda | H(\lambda) | \Psi_\lambda \rangle = 
\langle H(\lambda) \rangle_\lambda = E_\lambda[n_\lambda]$. 
$E_A[n_\lambda]$ and $E_B[n_\lambda]$ denote the energies of compound $A$ or $B$ 
evaluated using the wavefunctions (or density in the case of orbital free DFT) obtained at $\lambda$. 
Note that $n_{\lambda = 0} = n_A$, and $n_{\lambda=1} = n_B$.
Subtracting $E^{lin}(\lambda)$ and regrouping yields,
\bea
E[H(\lambda)] - E^{lin}(\lambda) & = & \left(E_A[n_\lambda] + \lambda(E_B[n_\lambda] - E_A[n_\lambda]\right) \nonumber\\
                              & & - \left(E_A[n_A] + \lambda (E_B[n_B]-E_A[n_A])\right)   \nonumber\\
&= & (1-\lambda) (E_A[n_\lambda]-E_A[n_A])  \nonumber\\
 & & + \lambda (E_B[n_\lambda]-E_B[n_B]), \nonumber\\
 & \ge & 0.
\eea
where the prefactors of the energy differences $\lambda, (1-\lambda) \ge$ 0 by definition, and where
$E_A[n_A] \le E_A[n_\lambda]$ and $E_B[n_B] \le E_B[n_\lambda]$ because of the variational principle.
Consequently, analogous inequalities will hold for any property for which there is a variational principle,
e.g.~also for the polarizability due to Pearson's maximum hardness principle~\cite{MaxHardnessPearson}.
This inequality is on display for the static polarizability, fractionally transmutating a hydrogen chloride molecule
into ammonia (Fig.~\ref{fig:HF}).
Similarly, potential energy inequalities in between different molecules were proposed by Mezey in the eighties~\cite{ConcavityMezey1985}.

Analytical first order derivatives of the energy as a function of some iso-electronic change in the
Hamiltonian can easily be calculated using the Hellmann-Feynman (HF) theorem~\cite{HF}, 
as proposed and demonstrated for HOMO eigenvalues in Ref.~\cite{anatole-jcp2009-2},
$\partial E/\partial \lambda  =  \langle \partial H/\partial \lambda \rangle_\lambda $. 
For a linearly interpolating Hamiltonian, such as in Eq.~(\ref{eq:Hamiltonian}), this leads to,
\bea
\frac{\partial E[H(\lambda)]}{\partial \lambda} & = & \left\langle H_B - H_A  \right\rangle_\lambda \;\;= \;\; 
\int d\fatr \; n_\lambda(\fatr) \times (v_B(\fatr)-v_A(\fatr)) \nonumber\\
& = & \left\langle H_B \right\rangle_\lambda - \left\langle  H_A \right\rangle_\lambda \;\;=\;\; E_B[n_\lambda] - E_A[n_\lambda].
\label{eq:HF}
\eea
The protonation energy, and its derivative, also feature in Fig.~\ref{fig:HF} for 
the same transmutational change, HCl $\rightarrow$ NH$_3$. 
As mentioned before, the use of pseudopotentials/valence electron densities fortunately renders straightforward
the application of the HF theorem according to Eq.~(\ref{eq:HF}) even for changes that involve
elements from differing rows in the periodic table.

Thermodynamic integration of $\partial E/\partial \lambda$ over $\lambda$ yields any properties related to free energy differences. 
In the case of compound design the approach is slightly different, we  
would like to expand the energy of a new compound $B$ in terms of a reference compound $A$ and its derivatives,
\bea
E_B & \approx & E_A[n_A] + \frac{\partial E_A[n_A]}{\partial \lambda} \Delta \lambda + \frac{1}{2} \frac{\partial^2 E_A[n_A]}{\partial \lambda^2}\Delta \lambda^2 + {\rm HOT}, \nonumber\\
\label{eq:Taylor}
\eea
HOT standing for higher order terms, and $\Delta \lambda = 1$. 
Unfortunately, when making predictions with a linearly interpolated Hamiltonian, 
the first order derivative term according to Eq.~(\ref{eq:HF}) is not necessarily predictive~\cite{anatole-jcp2009-2}.
Unfortunately, the inclusion of higher order derivatives in Eq.~(\ref{eq:Taylor}) might not only improve the prediction,
as found for statistical mechanical averages~\cite{HigherOrderAlchemicalDerivatives_SmithGunsteren1994},
but it also requires the evaluation of the perturbed wavefunction, 
e.g.~through the use of linear response theory~\cite{apdsmp,anatole-jcp2005}, 
thereby defying the original purpose of predicting a new compound's energy without having to calculate its wave function. 
Nevertheless, for external potentials this has been carried out within conceptual DFT~\cite{2ndOrderDerivsWrtVext_DeProftAyersGeerlings},
and very recently even analytically~\cite{YangCohenGeerlingsAnalyticalFukui2012}.

In order to improve the predictive power of the first order term in Eq.~(\ref{eq:Taylor}), 
an empirical correction has been introduced that ``linearizes'' the energy through a global yet 
non-linear Hamiltonian, $H(\lambda) = H_A + f_{AB}(\lambda)(H_B-H_A)$~\cite{anatole-jcp2009-2}.
If we assume $f(\lambda)$ to be a second order polynomial in $\lambda$, two coefficients
are determined by the boundary conditions that $H(\lambda=0) = 0$, and $H(\lambda=1) = 1$, leaving one additional degree of freedom. 
We can obtain the remaining degree of freedom as a parameter from 
an arbitrary second iso-electronic compound pair, $CD$, such that the energy becomes linear in $\lambda$.
The resulting expansion up to first order in Eq.~(\ref{eq:Taylor}) then becomes,
\bea
E_A & \approx & E_A[n_A] + C^{ref}_{CD} \frac{\partial E_A[n_A]}{\partial \lambda} \Delta \lambda.
\label{eq:predictref}
\eea
Here, $C^{ref}_{CD}$ is the ratio between the energy difference and the HF derivative of
the additional reference compound pair, $C$ and $D$ (its Hamiltonian being linearly interpolated),
and $\partial_\lambda E_A[n_A]$ is determined according to Eq.~(\ref{eq:HF}) as $E_B[n_A]-E_A[n_A]$. 
This bears resemblance to a long tradition in physical chemistry, 
namely the use of reference compounds for electrode potentials or enthalpies.

The idea to use alternative, non-linear, interpolations is not new within the molecular mechanics research.
In the context of electronic structure theory non-linear alchemical paths were also explored 
for chemical binding~\cite{ArianaAlchemy2006}, and nuclear quantum effects~\cite{alejandro-jctc2011}.
Various open questions deserve further investigation,
such as transferability and choice of reference coefficients, 
iso-electronic changes using valence electrons only versus all electron description,
non-iso-electronic changes, necessary accuracy when providing the input of target compound $B$, i.e. also its geometry, 
ionic forces of $B$, etc. These answers are likely to depend on systems and properties.

\subsection{Control of ligand binding}
\begin{figure}
\includegraphics[scale=0.26, angle=0]{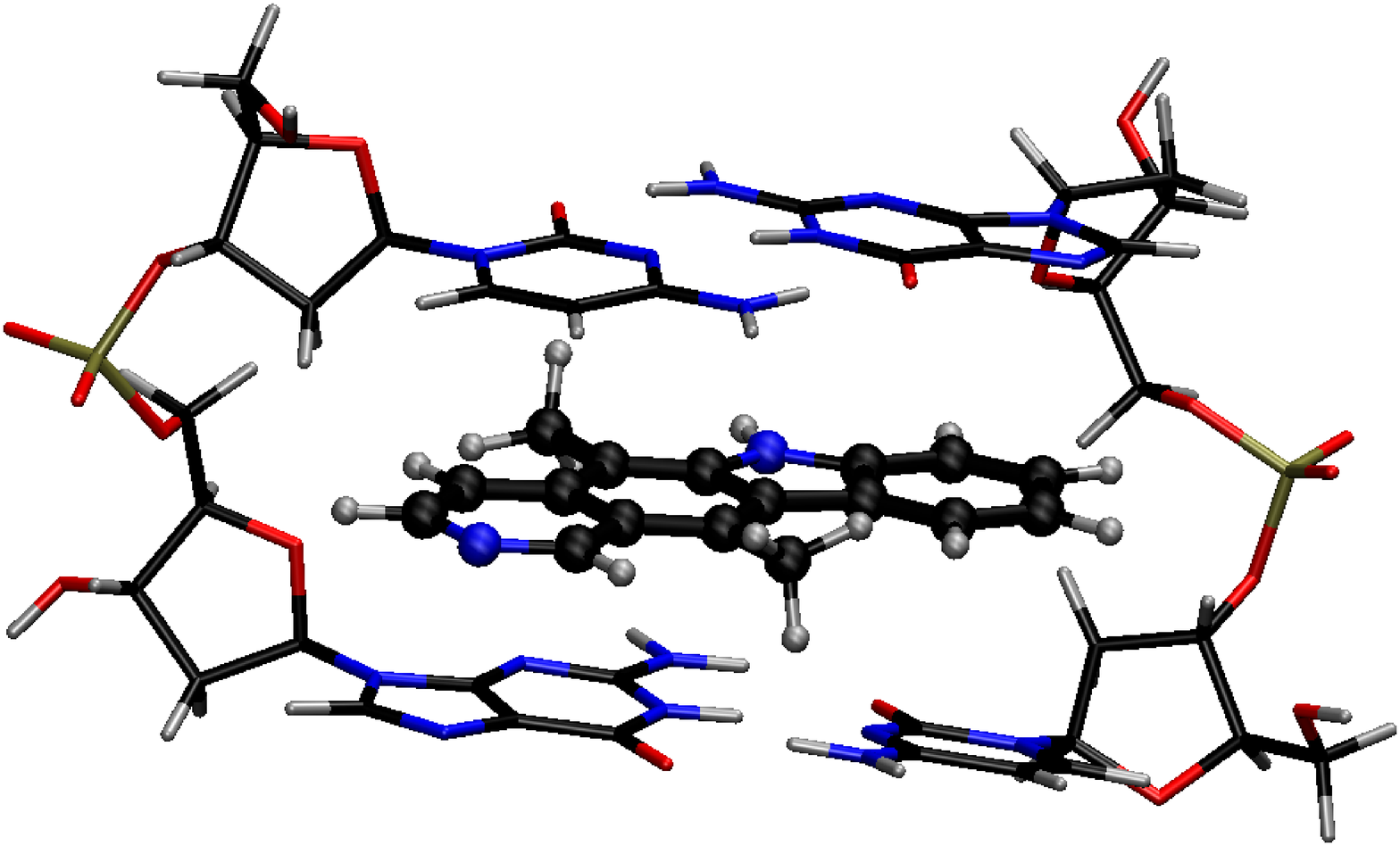}
\includegraphics[scale=0.64, angle=0]{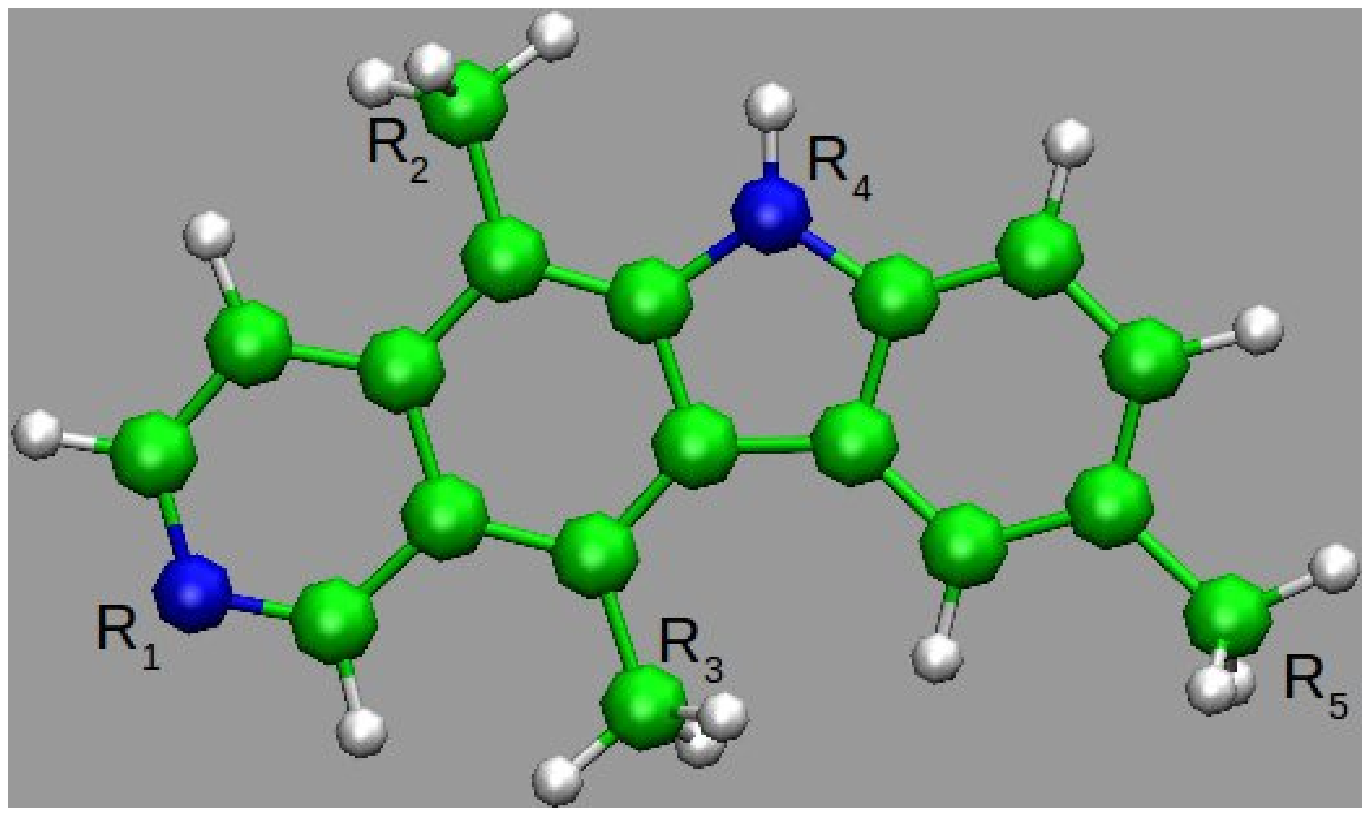}
\caption{
\label{fig:elli1}
TOP: Cluster model of drug intercalated in between two Watson-Crick base-pairs connected by
sugar puckers and phosphate groups.
BOTTOM: Neutral ``wild-type'' ellipticine, R$_i$ denote sites of groups permitted to mutate (see TAB.~\ref{tab:elli}).
}
\end{figure}

In this section, we exemplify the use of the reference coefficients [Eq.~(\ref{eq:predictref})]
for increasing the predictive power of the HF derivatives of linearly interpolated Hamiltonians. 
We refer to state-of-the-art van der Waals corrected DFT~\cite{anatole-prl2004,anatole-prb2007} 
to accurately estimate interaction energies with binding targets across CCS. 
We will consider a small yet illustrative set of mutants of the ellipticine molecule.
Ellipticine is a naturally occurring anti-cancer drug with various binding targets.
As also illustrated in Fig.~(\ref{fig:elli1}), its dominant mode of binding to DNA is intercalation. 
Structural data as well as studies on drug analogues are 
readily available~\cite{EllipticineMutants_Frei2004,EllipticineBenzoxazinicAnalogues}.
We will probe the versatility of the linearizing scheme for controlling ellipticine-derivatives/DNA binding~\cite{EllipticineXray_2005}.
Clearly, for the eventual control of ligand binding the property of interest is not the potential energy of interaction
but rather the free energies of binding: Solvation or entropic contributions can be crucial, 
as is well known in general~\cite{BiomolecularInteractions_Stahl2010}, 
and in the particular case of ellipticine~\cite{EllipticineEntropy_Hobza2010}. 
For example, Tidor~\cite{Lambdaspacetidor}, and Oostenbrink and van Gunsteren~\cite{OostenbrinkReferencCMPD2005,OostenbrinkHostGuest2008}
have carried out similar work in the sense of interpolating ligand candidates, 
by calculating free energies of binding, and using molecular force-fields.
However, here we will focus on the potential energy of interaction. 
Subsequent work in the future can deal with the inclusion of thermal and solvent effects for instance using {\em ab initio} molecular 
dynamics techniques~\cite{AIMD_PNAS_TUCKERMAN2005} in conjunction with QM/MM~\cite{laio-qmmm} calculations.
Moreover, even at the mere potential energy electronic structure level of theory, the accurate quantification and control
of intercalated ellipticine derivatives is challenging: vdW forces dominate the binding. 
Recent studies have already explored the binding of ellipticine and 
how its vdW forces can be accounted for at the employed electronic 
structure level~\cite{anatole-jpcb2007,anatole-jcp2010,mbd_PNAS2012}. 

Let us consider the intercalation energy for the complex depicted in Fig.~(\ref{fig:elli1}) 
for mutations at the five sites indicated in the bottom panel.
In analogy to protein or DNA sequences, an (arbitrary) relevant subspace of CCS is defined in
TAB.~(\ref{tab:alphabet}) as a matrix that corresponds to an alphabet of
iso-electronic (in valence electron number) functional groups at each of the selected sites.
Note that variation in molecular combinations of letters of this alphabet are capable to not only revert dipole-moments,
they can also act either as hydrogen bond acceptors (lone pair in OH/Cl) or donors (NH$_2$, proton in OH).
Clearly, the alphabet can easily be extended to accommodate further effects, for example with electron donating/withdrawing or 
hyperconjugating groups etc. Conformational degrees of freedom can be encoded explicitly,
as it is done for the hydroxyl groups in TAB.~(\ref{tab:alphabet}).

Within this restricted CCS, any given molecule is represented by the sequence of functional groups distributed over the five sites.
For example, the ``wild-type'' ellipticine in Fig.~(\ref{fig:elli1}) would correspond to (21121),
{\em i.e.} 2 for N at R$_1$, 1 for CH$_3$ at R$_2$, R$_3$, and R$_5$, and 2 for NH$_2$ at the R$_4$.
Let us exemplify a DFT+vdW based prediction of the binding energy of another mutant:
$(21121)$ is predicted to bind to the DNA cluster in Fig.~\ref{fig:elli1} with $E_{21121}$ = $-38.5$ kcal/mol~\cite{anatole-jpcb2007}.
For predicting the single point mutation (21121)$\rightarrow$(21125) (changing CH$_3$ into F at R$_5$),
one would have to predict a target value of $E_{21125}$ = $-36.9$ kcal/mol.
The derivative based prediction according to first order term in Eq.~(\ref{eq:Taylor}) is calculated to be,
$E_{21125} \approx E_{21121}$ + $\partial_\lambda E_{21121}$ = $-38.5$ + 1.4 = $-37.1$ kcal/mol.
Inclusion of reference coefficient [Eq.~(\ref{eq:predictref})], and using compound pair (11121)/(11125) as a reference,
yields
$E_{21125} \approx E_{21121}$ + $C_{ref} \times \partial_\lambda E_{21121}$ = $-38.5$ + 1.3$\times$1.4 = $-36.7$ kcal/mol.

\begin{table}
\caption{
Exemplary alphabet for mutants of ellipticine as oriented in Fig.~\ref{fig:elli1}, defining a CCS with 4x6$^4$ = 5184 molecules. 
Highlighted in red are all functional groups whose mutations have been considered. Predictions are displayed in Fig.~\ref{fig:elli2}.
The ``wild-type'' ellipticine drug is encoded as (21121) with three functional groups coming from the first column, 
and the functional groups at site R$_1$  and R$_4$ coming from the second column. 
}
\label{tab:elli}
\begin{tabular}{|l|c|c|c|c|c|c|} \hline
site vs.~group &  1 &  2 & 3 & 4  & 5& 6 \\ \hline
R$_1$ &  {\color{red} CH}  & {\color{red} N} & SiH & P  & - & -\\ \hline
R$_2$ &  CH$_3$ &  NH$_2$ & OH$^{\rm left}$ & OH$^{\rm right}$  & F & Cl  \\ \hline
R$_3$ &  CH$_3$ &  NH$_2$ & OH$^{\rm left}$ & OH$^{\rm right}$  & F & Cl  \\ \hline
R$_4$ &  {\color{red} CH$_2$} & {\color{red} NH} & O & SiH$_2$ & PH  & S   \\ \hline
R$_5$ &  {\color{red} CH$_3$} & NH$_2$ & OH$^{\rm left}$ & OH$^{\rm right}$  & {\color{red} F} & Cl  \\ \hline
\end{tabular}
\label{tab:alphabet}
\end{table}

\begin{figure}
\includegraphics[scale=0.4, angle=0]{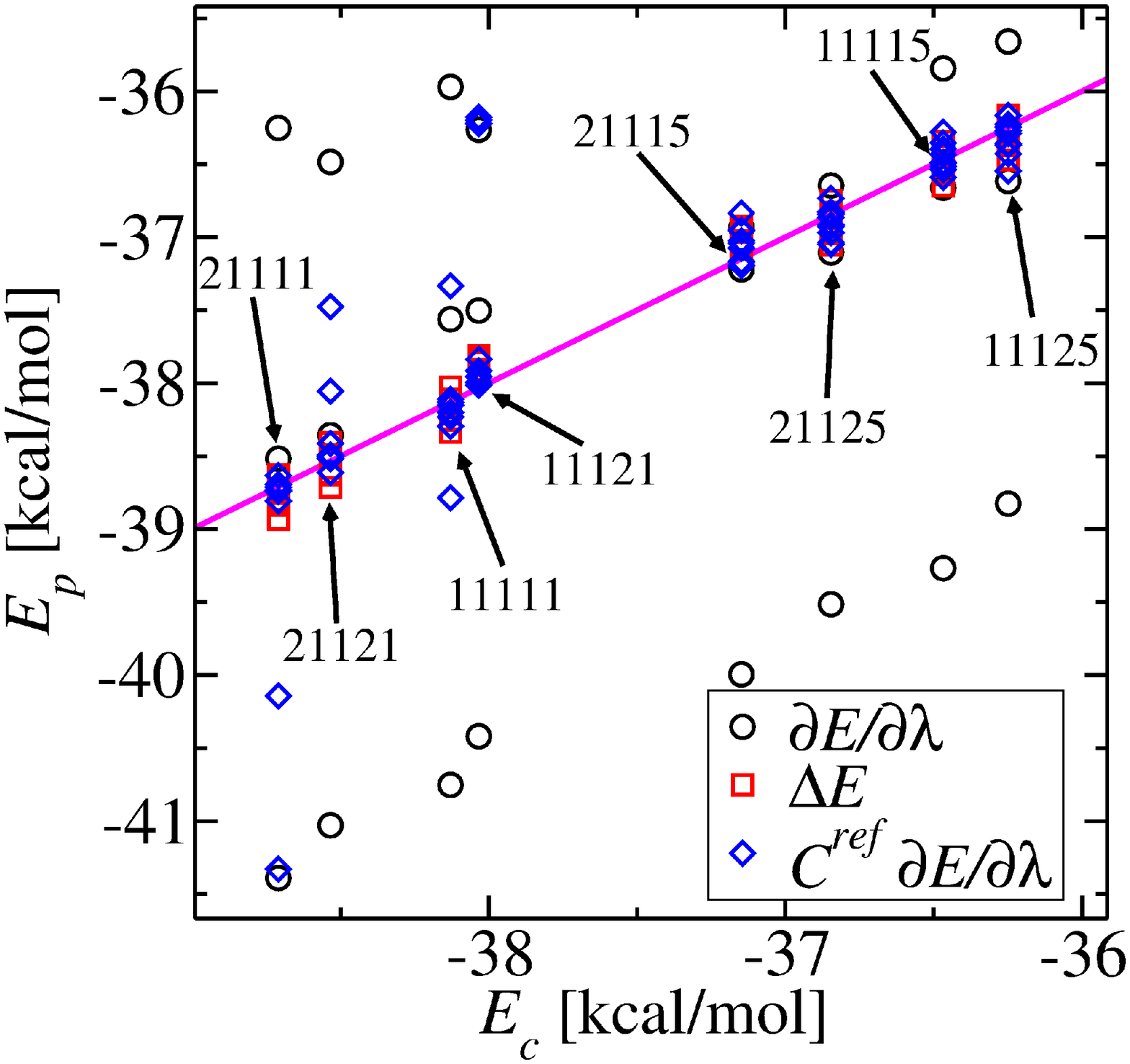}
\includegraphics[scale=0.33, angle=0]{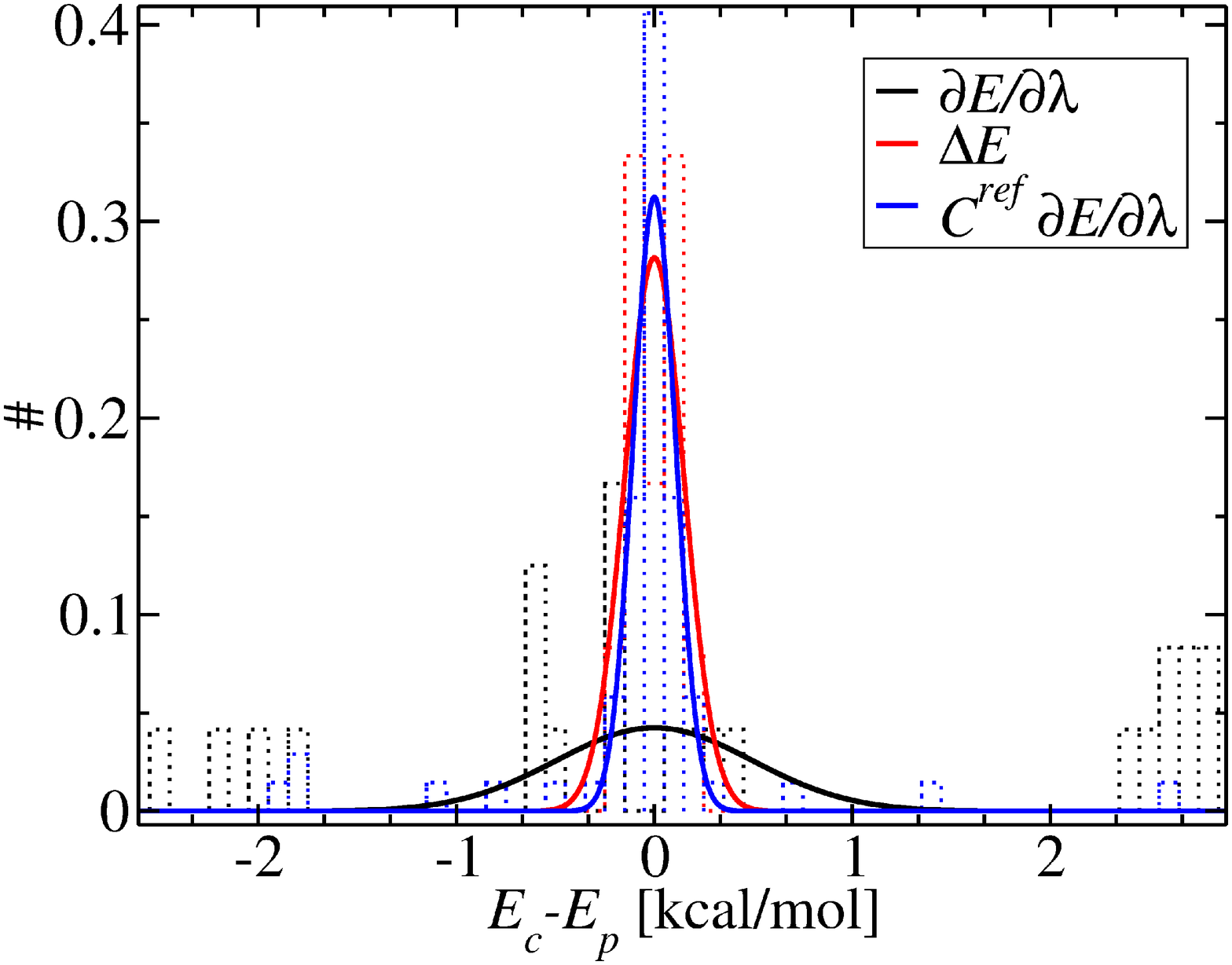}
\caption{
\label{fig:elli2}
TOP: Correlation for eight compounds from alphabet in TAB.~\ref{tab:alphabet}.
Predictions made using first order derivatives only ($\partial E/\partial \lambda$, Eq.~(\ref{eq:Taylor})),
energy difference of reference compound pairs ($\Delta E$, Eq.~(\ref{eq:deint})),
or $C^{ref} \partial E/\partial \lambda$, Eq.~(\ref{eq:predictref}).
BOTTOM: Normalized histogram and corresponding normal distribution of error over 72 predictions,
$\Delta_{\rm c-p}$E$^{\rm int} =$ calculated E$^{int}$ - predicted E$^{int}$.
}
\end{figure}

In order to gain a more representative idea of the predictive power of this method,
Fig.~\ref{fig:elli2} features the outcome for a small subspace of the CCS highlighted in red in TAB.~\ref{tab:alphabet}:
Eight compounds have been considered involving permutations at R$_1$, R$_4$, and R$_5$, each with two possible functional groups.
Predictions based on all the possible derivatives among these compounds, with
and without reference coefficients (as obtained from compound pairs not involved in the transmutation),
are compared to calculated binding energies.
Despite the several outliers that deviate substantially, the use of reference compounds dramatically improves the overall prediction.

For comparison, we also include predictions based on the additive assumptions that the influence
of the rest of the molecule cancels when considering an interpolation for the same pair of functional groups. 
Specifically, we estimate the binding energy of B 
simply by adding the difference in binding energy of a reference compound pair, $CD$, to the binding energy of $A$,
\begin{eqnarray}
E_{\rm B} & \approx & E_{\rm A} + \Delta E \;\; = \;\; E_{\rm A} +  (E_{\rm D} - E_{\rm C})
\label{eq:deint}
\end{eqnarray}
As shown in Fig.~\ref{fig:elli2}, also this prediction yields remarkable good correlation---with less pronounced outliers.
Analysis of the distribution of errors, however,  suggests that in spite of the outliers
the normal distribution of predictions around the ideal correlation is superior for 
predictions made with the product of derivative and reference coefficient (Bottom of Fig.~\ref{fig:elli2}).

\subsection{Win a prize}
The numerical illustrations in the previous section, as well as in Ref.~\cite{anatole-jcp2009-2}, 
suggest that efforts to linearize the property through use of alternative, non-linear interpolations of 
the Hamiltonians are worthwhile. Strictly speaking, however, due to the use of reference compound pairs, 
the aforementioned interpolation constitutes no longer a first principles but rather an empirical and heuristic {\em Ansatz}. 
What is needed instead, is an {\em ab initio} interpolating procedure that linearizes the energy (or other properties)
in order parameter, such that the first order Taylor expansion based on the HF derivative is sufficiently 
accurate to predict properties of other compounds~\cite{AbInitioDefinitionByKieronBurke}.

Inspired by Erd\"os' habit to offer cash awards for solutions to outstanding mathematical problems, 
the author has thus decided to offer the equivalent of an ounce of gold to the first person who presents an {\em ab initio} 
solution to this problem. 
Specifically, the challenge reads: Find---or show non-existence of---an {\em ab initio}, 
i.e. valid for {\em any} external potentials, interpolating transform $f_{AB}(\lambda)$ for which 
two different but iso-electronic molecular Hamiltonians with energies $E_A$ and $E_B$ 
interconvert such that the electronic ground state potential energy $E=\langle H(f(\lambda))\rangle$, 
is linear in order parameter $\lambda$, and that consequently the HF derivative is given by,
\bea
\frac{\partial E(\lambda)}{\partial \lambda}\Bigg|_{\lambda} & = & \left\langle \frac{\partial H(f(\lambda))}{\partial \lambda} \right\rangle_{\lambda} \nonumber\\
& = & E_B - E_A.
\eea
Here, 0 $\le \lambda \le$ 1, and $E(\lambda = 0) = \langle H(f(\lambda=0)) \rangle = \langle H_A\rangle = E_A$,
and $E(\lambda = 1) = \langle H(f(\lambda=1)) \rangle = \langle H_B\rangle = E_B$.
Further details can be found in footnote~\cite{vonLilienfeldPrize}.

We can exemplify the challenge by solving it for the non-relativistic hydrogen-like single atom with only one electron. 
In this case, $E(\lambda) = a Z(\lambda)^2$, where $a$ is a constant.
and the nuclear charge $Z$ is a function of interpolating parameter $\lambda$~\footnote{We assume the reduced mass to equate the mass of the electron}.
For an interpolation linear in the Hamiltonian, $Z(\lambda) = Z_A + \lambda (Z_B-Z_A)$, 
and the energy is therefore clearly quadratic in $\lambda$.
The desired behavior of a linearized energy would be, 
\bea
E^{lin}(\lambda) & = & E(Z_A) + \lambda(E(Z_B) - E(Z_A)) \nonumber\\
& = &  a (Z_A^2 + \lambda(Z_B^2 - Z_A^2)).
\eea
Equating this to $a Z(\lambda)^2$ and solving for $Z(\lambda)$ yields the corresponding interpolating function:
\bea
Z(\lambda) & = & \sqrt{Z_A^2 + \lambda (Z_B^2 - Z_A^2)}. \label{eq:Interpolation}
\eea
As suggested above in the challenge, we can test this interpolation to confirm if indeed we find the desired slope for the linearized energy, $E_B-E_A$.
Application of the chain rule, and insertion and differentiation of Eq.~(\ref{eq:Interpolation}) confirms the expected result,
\bea
\frac{\partial E}{\partial \lambda}\Bigg|_{\lambda} & = & \frac{\partial E}{\partial Z}\frac{\partial Z(\lambda)}{\partial \lambda}\Bigg|_{\lambda}
\;\;= \;\; 2 a Z(\lambda) \frac{Z_B^2 - Z_A^2}{2 \sqrt{Z_A^2 + \lambda (Z_B^2 - Z_A^2)}}\Bigg|_{\lambda}, \nonumber\\
& = & a(Z_B^2 - Z_A^2) \;\;\equiv\;\; E_B - E_A.
\eea
As such Eq.~(\ref{eq:Interpolation}) linearizes the energy in $\lambda$. 
The challenge of the prize consists of finding an analogous expression for molecules, 
i.e.~a spatially resolved and $\lambda$ dependent distribution of nuclear interpolations, $\{Z(\fatR_I,\lambda)\}$, 
that drive all atoms in compound $A$ to atoms in compound $B$ while linearizing the potential energy.

Note that a naive extension of Eq.~(\ref{eq:Interpolation}) to assemblies of atoms, 
\bea
\frac{\partial E}{\partial \lambda} & = & \sum_{I\in A,B} \frac{\partial E}{\partial Z_I} \frac{\partial Z(\lambda)}{\partial \lambda} 
\;\; = \;\; \sum_{I\in A,B} \mu_p(\fatR_I) \frac{\partial Z(\fatR_I,\lambda)}{\partial \lambda} \nonumber\\
\eea
does not constitute a practical approximate solution to the challenge. 
$\mu_p(\fatR_I)$ denotes the ``alchemical'' potential mentioned above
which corresponds to the electrostatic potential at $\fatR_I$ without the repulsion due to $Z_I$.
$\mu_p$ will not necessarily cancel the square root term in the denominator of 
the derivative in Eq.~(\ref{eq:Interpolation}), which consequently
diverges if $\lambda$ and $Z_A(\fatR_I)$ equal 0.

\section{Statistical methods}
\label{sec:ML}
\subsection{Inductive reasoning from first principles}
Within statistical mechanics the numerical prediction of macroscopic observables from atomistic simulation requires repeatedly 
calculating microscopic states, using electronic structure theory, atomistic or coarse-grained force fields, 
and averaging in an appropriate ensemble.
Philosophically speaking, the exercise of performing such computational ``experiments'' is an application of 
{\em deductive} reasoning to increase knowledge.
But also when exploring CCS in terms of ensembles of potential energy hypersurfaces by repeatedly solving 
SE for $N$ different compounds deductive reasoning is at work.
Since the size of CCS is prohibitively large, its exhaustive exploration through screening with SE is impossible. 
While some interpolating $\lambda$ schemes use statistical mechanics for a preselected set of 
compounds~\cite{Lambdaspacetidor,OostenbrinkReferencCMPD2005}, 
 a rigorous way to systematically and generally gain quantitative insights is desirable.
This task can be accomplished through the application of {\em inductive} reasoning.

Historically, the role of inductive reasoning for chemistry is considerable, 
Mendeleev's table, the Hammett equation, or Pettifor's structure maps~\cite{Hammett_cr1935,Hammett_jacs1937,StructureMaps_Pettifor1986} are  
all based on inferred phenomenological relationships. 
Further examples include widely spread rules and notions of chemistry, such as the chemical bond, atomic charges, or aromaticity.
While popular and useful to the experimental chemist conventional quantum chemistry, based on deductive reasoning, 
is still struggling to account for these notions~\cite{FuzzyChemicalConcepts_Corminboeuf2012}.
Recent advances in statistical data analysis methods~\cite{HasTibFri01,MLbook,htf2009,MueMikRaeTsuSch01}
and applications in other areas of science and engineering, such as 
searching the internet, automated locomotion (self-driving cars), algorithmic trading,
or brain-computer interfaces, strongly suggest that they will also play an increasingly important role in chemistry.
Examples of first efforts to quantitatively infer laws for atomistic simulations include 
``Learning On The Fly''~\cite{lotf2004}, or ``force-matching''~\cite{PatrickForceMatchingJCTC2007,VoorhisForceMatchingJCTC2010}.
More sophisticated statistical learning methods have been applied to the training of
exchange correlation functionals in density-functional theory~\cite{NN4B3LYP_Chen2003,NN4B3LYP_Chen2004},
or to parameterizing interatomic force fields~\cite{PotentialEnergyFit_Bowman2003,Neuralnetworks_Scheffler2004,Neuralnetworks_BehlerParrinello2007,MachineLearningWaterPotential_Handley2008,Neuralnetworks_Behler2008,b2011d}.
Support vector machines have been shown to quantify basis-set incompleteness~\cite{SVM4CBS_Lomakina2011}.
Gaussian kernel based machine learning (ML) for very accurate reactive force-fields was introduced by Bartok et al.~\cite{bpkc2010}.
Contributions by Curtarolo, Hautier, and Ceder combine data-mining with 
mean-field electronic structure theory~\cite{CurtaroloPRL2003,CurtaroloCMS2010,MachineLearningHautierCeder2010}.
Even the learning of reorganization energies that enter Marcus charge transfer rates are promising~\cite{RatnerJACS2005,anatole-MilindDenis2011}.
Very recently, kernel based ML models have also delivered promising results for learning electron density functionals~\cite{ML4Kieron2012},
or transition state theory dividing surfaces that determine reaction rates~\cite{ML4Graeme2012}.
Bayesian error estimates and cross-validation methods have also been applied to the development of
exchange-correlation models with controlled transferability~\cite{Bayesian4DFT_Jacobsen2012}.

Within the bioinformatics and cheminformatics communities the development of quantitative structure property
relationships (QSPRs) has a long tradition. 
QSPRs, relying on similar statistical frameworks (ML, cross-validated training, principal component analysis, etc.),
deliberately attempt to circumvent solving the underlying laws of physics by directly correlating system parameters 
(descriptors) with macroscopic properties of interest.
Conventionally, QSPRs are based on descriptors that explicitly forsake atomic resolution in the first principles sense.
A large variety of such QSPR descriptors for various properties has been 
proposed~\cite{SchneiderReview2010,WienerDescriptors,TodeschiniConsonniHandbookDescriptor,DescriptroOverviewMeringer2005}.
Two such descriptors, the molecular signature by Faulon and coworkers~\cite{SignatureFaulon2003}
and a combination of HOMO eigenvalues of charged and neutral species, have recently yielded promising results for the QSPR modeling of
a first principles property, the reorganization energy, in the CCS of polycyclic aromatic hydrocarbons (PAHs)~\cite{anatole-MilindDenis2011}.
PAHs are used in discotic liquid crystals which self-assemble into columnar liquid crystal structures,
implying their usefulness for organic photo-voltaic applications~\cite{AndrienkoKremerMullen2009}.

In this section we will discuss the application of {\em ab initio} statistical learning approaches 
to previously obtained first principles data for $N$ compounds. 
Merely based on the data, QSPRs can be inferred, i.e. ``learned'', 
and subsequently be used to avoid the cumbersome task of having to explicitly model all 
the underlying physical degrees of freedom of electrons and nuclei. 
As such, ML estimates solutions of SE for a new, i.e. ``unseen'', molecule $B$ simply 
by evaluating an analytical expression $E^{est}(B)$ that (explicitly or implicitly) 
encodes the data of $N$ other molecules.
Obviously, any such inferred relationships are inherently limited in accuracy by the quality of the data used for training. 

\subsection{Machine learning in CCS: The quantum machine}
Recently, a kernel ridge regression approach to learn DFT atomization energies across CCS has been introduced~\cite{RuppPRL2012}.
Unlike ordinary QSPR approaches, this ML model is free of any heuristics. 
It exactly encodes the supervised learning problem posed by SE, 
i.e.~instead of finding the wavefunction $\Psi$ which maps the system's Hamiltonian to its energy,
$H(\{Z_I, \fatR_I\}) \stackrel{\rm \Psi}{\longmapsto} E$, 
it directly maps system to energy based on $N$ examples given.
In the limit of converged $N$, i.e.~sufficiently dense system coverage, 
the ML model is therefore a formally exact inductive equivalent to the deductive solution 
of SE through use of approximate wave-functions (such as separability of nuclear and electronic wavefunction or single slater determinants),
Hamiltonians (such as certain exchange-correlation potentials), and self-consistent field procedure to minimize the energy.
In Ref.~\cite{RuppPRL2012} numerical evidence is given for this idea. 
Specifically, for a diverse set of organic molecules, one can show that a ML model can be used instead,
$\{Z_I, \fatR_I\} \stackrel{\rm ML}{\longmapsto} E$.
After training, solutions to SE can be inferred for out-of-sample, i.e. ``unseen'', 
compounds that differ either in geometry or in composition or in both. 
The evaluation of an estimate is ordinarily negligible in terms of computational cost, 
i.e.~milli seconds instead of hours on a conventional CPU,
while yielding an accuracy competitive with the deductive approaches of modern electronic structure theory.
As within any inductive approach, the accuracy is limited by the domain of applicability  
as defined by the data used for training, i.e.~robust results can only be expected 
in interpolating regimes with sufficient coverage. 
Within the Gaussian kernel model, the energy of a query molecule $\fatM_A$ is given as a sum 
over $N$ molecules in the training set, 
\bea
E^{est}(\fatM_A) & = & \sum_{i=1}^N \alpha_i \;\; e^{-\frac{d(\fatM_A,\fatM_i)^2}{2 \sigma^2}}.
\label{eq:MLenergy}
\eea
Each training molecule $i$ contributes to the energy according to its specific weight $\alpha_i$, 
scaled by a Gaussian in its distance to $\fatM_A$, $d(\fatM_A,\fatM_i)$.
For given length-scale $\sigma$ and regularization parameter $\lambda$,
$\{\alpha_i\}$ are obtained by solving the regression problem,
\bea
\underset{{\bm \alpha}}{\rm min} && \sum_i \bigl( E^{est}(\fatM_i) - E^{ref}_i \bigr)^2 + \lambda \sum_i \alpha_i^2. \label{eq:krr}
\eea
$\sigma$ and $\lambda$ are hyperparameters. 
This regularized model limits the norm of regression coefficients, $\{ \alpha_i \}$,
thereby improving the transferability of the model to new compounds.
All regression coefficients and hyper-parameters are determined by cross-validation
on data stratified training sets~\cite{htf2009,MueMikRaeTsuSch01}.

\begin{figure} 
\centering
\includegraphics[scale=0.31, angle=0]{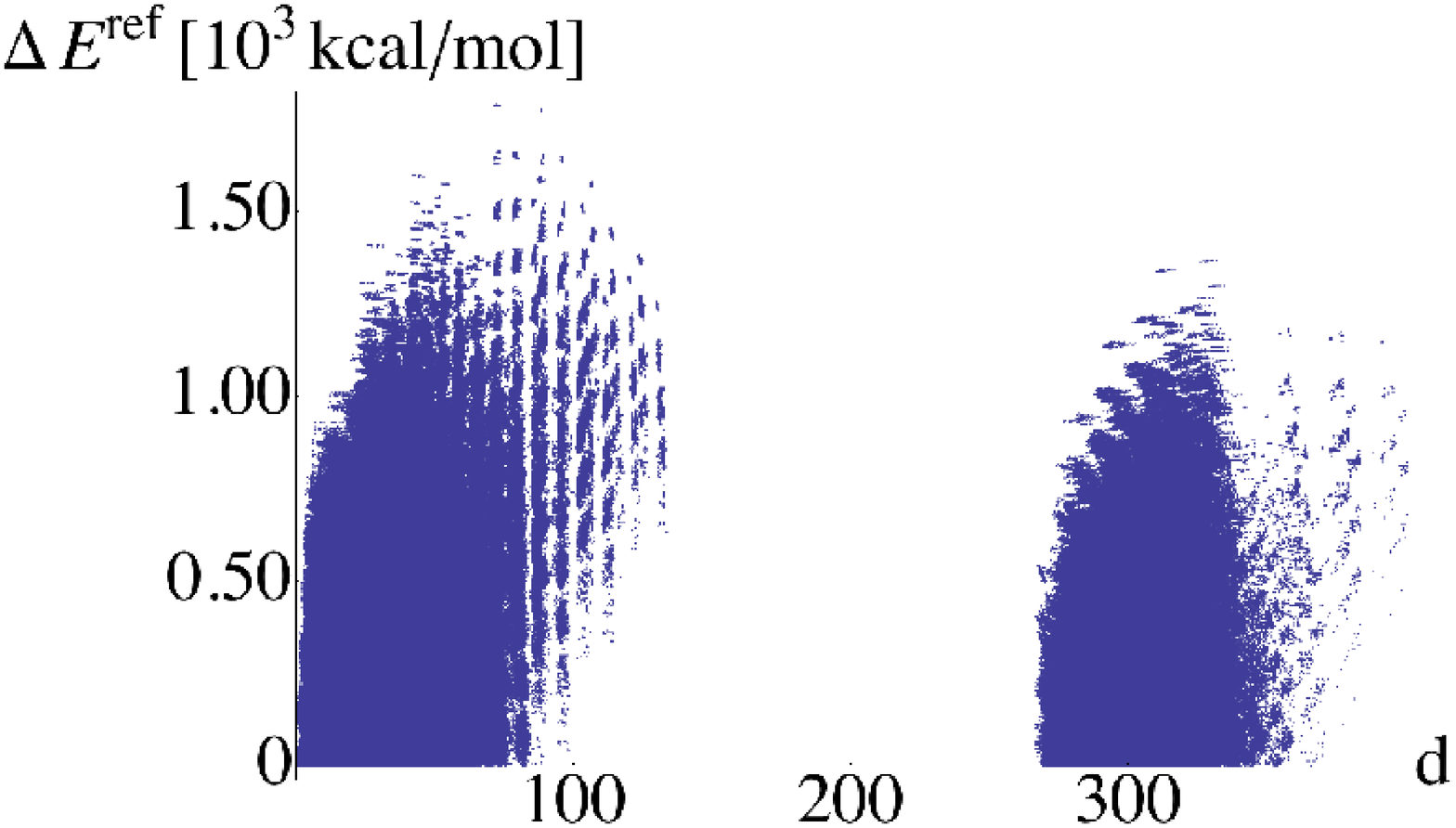}
\includegraphics[scale=0.3, angle=270]{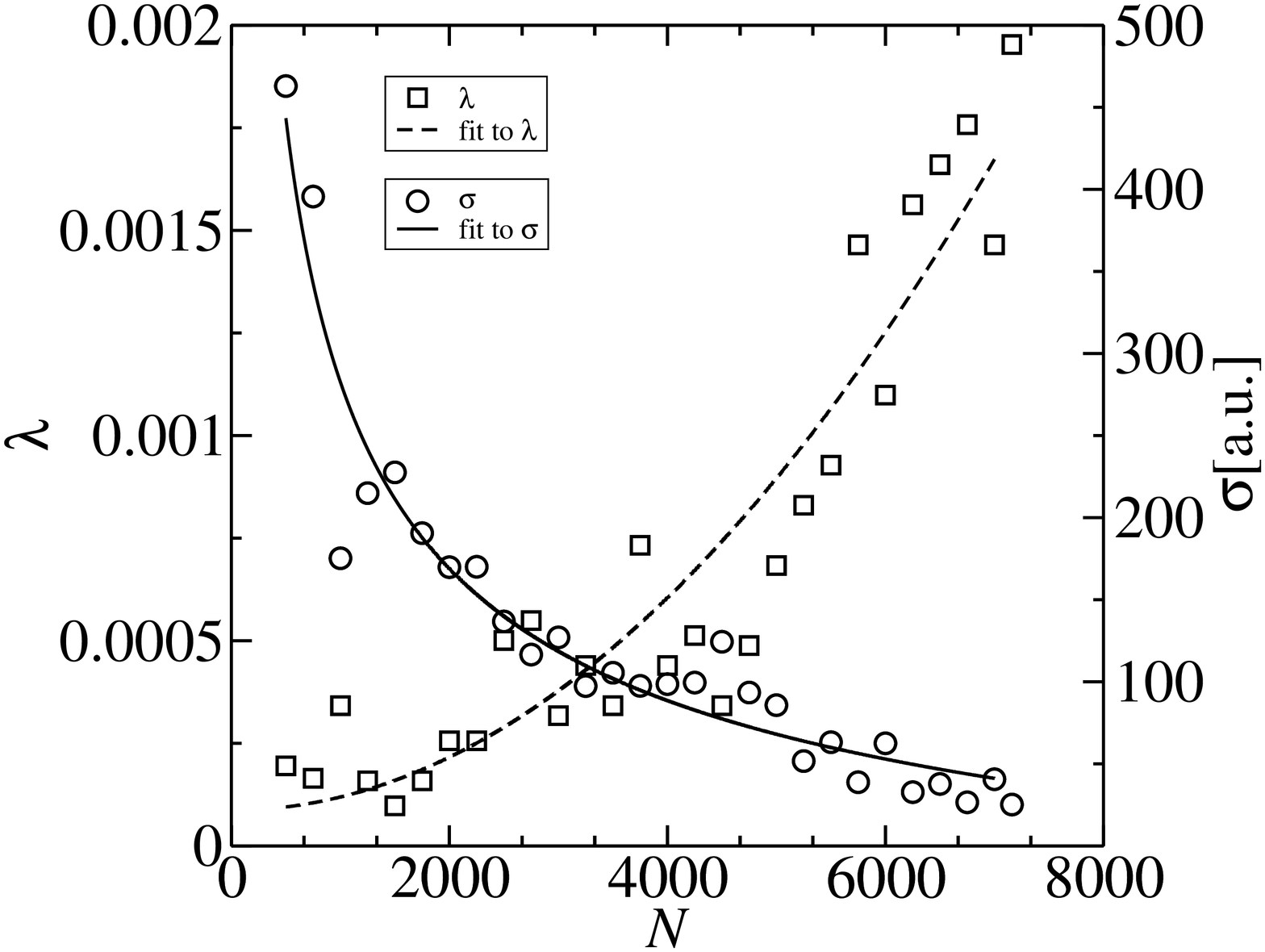}
\caption{
TOP: Distance distribution in GDB-13 for the 7000 smallest molecules, $\Delta E^{\rm ref} = |E_i - E_j|$ versus $d(\fatM_i,\fatM_j)$.
BOTTOM: $N$ dependence of $\sigma$ and $\lambda$.
}
\label{fig:locality}
\end{figure}

So far this model has been trained and validated only in its most rudimentary form for
atomization energies of a small set of interesting compounds.
Specifically, molecular atomization energies at the hybrid DFT level of theory~\cite{HK,KS,becke3,PBE0,PBE01}
have been used for training on up to $N \approx 7000$ molecules from the GDB data base~\cite{ReymondChemicalUniverse3}
(see Fig.~\ref{fig:locality} for an illustration), for which mean absolute errors of less than 10~kcal/mol have been obtained.
The choice of hybrid DFT is motivated by relatively small errors ($<$ 5~kcal/mol) for thermo-chemistry
data that includes molecular atomization energies~\cite{DFT4G3_Truhlar2003}.
While 10 kcal/mol is still far from ``chemical accuracy'' ($\approx$ 1 kcal/mol), 
more recent progress has not only led to atomization errors with less than 3 kcal/mol accuracy~\cite{Montavon_NIPS2012},
but also includes other electronic properties, such as frontier eigenvalues, polarizability, and excitation energies~\cite{Montavon2012}.

An appealing advantage of analytical models, independent if obtained from physical insight or statistical
regression, is their amenability to physical analysis. 
For example, unlike electronic structure methods, otherwise ill-defined concepts such as
distance/neighborhood/similarity in CCS can now be quantified within the ``world'' of the ML model.
Specifically, Eq.~(\ref{eq:MLenergy}) gives the energy of a query molecule $\fatM_A$ as an expansion in compound space spanned
by reference molecules $\{\fatM_i\}$: The regression weights $\{\alpha_i\}$ are scaled by the similarity
between query and reference compound as measured by a Gaussian of the distance.
Hence, $\alpha_i$ assigns a positive or negative weight to molecule $i$.
Within the compound space used as reference, molecules therefore can be ranked according to their $|\alpha|$.
However, since $\{\alpha_i\}$ are regression coefficients in a non-linear model, i.e.~after a non-linear transformation
of the training data, the resulting energy contributions are specific to the employed training set without
general implications for other properties or regions of compound space.
The locality of the model is measured by $\sigma$, enabling the definition of a critical distance of locality, $d_c$,
i.e.~only if $d(\fatM_A,\fatM_i) \le d_c$ will $\fatM_i$ contribute to the energy of $\fatM_A$ more than some
threshold energy $E_c$.
Rearranging summands in Eq.~(\ref{eq:MLenergy}) leads to $d_c(\fatM_A,\fatM_i) = \sigma \sqrt{2 \ln[\alpha_i/E_c]}$.
For atomization energies, and the chemical space considered in Ref.~\cite{RuppPRL2012},
i.e.~with a critical distance $\leq$ 400 Bohr (see TOP of Fig.~\ref{fig:locality})
the ML results suggest that the model becomes local when $\sigma \leq$ 60 Bohr, for the average $\alpha$, and for $E_c$ = 1 kcal/mol.
Such $\sigma$ values are achieved when the number of molecules in training set $N$ exceeds $\sim$5000.
In other words, for $N \leq$ 5000, the model is global, i.e.~all reference compounds contribute with more than 1 kcal/mol 
to any prediction made.
See BOTTOM of Fig.~\ref{fig:locality} for the $N$ dependence of $\sigma$ and $\lambda$.

\subsection{Coulomb matrix descriptor}
To represent compounds, a wide variety of ``descriptors'' is in use by statistical methods for chem- and bio-informatics
applications~\cite{DescriptroOverviewMeringer2005,SchneiderReview2010,SignatureFaulon2003,WienerDescriptors,TodeschiniConsonniHandbookDescriptor}.
The descriptor introduced by Rupp et al.~\cite{RuppPRL2012} is based solely on coordinates and nuclear charges, 
and dubbed ``Coulomb-matrix'', $\fatM$, a symmetric square matrix of $N_I \times N_I$ dimensions,
\bea
M_{IJ}  =
\begin{cases}
0.5  Z_I^{2.4} & \forall \;\; I = J,\\
   \frac{Z_IZ_J}{|\fatR_I - \fatR_J|} & \forall \;\; I \ne J.
\end{cases}
\label{eq:matrix}
\eea

The diagonal elements, $E(Z_I) \approx 0.5 Z^{2.4}_I$, correspond to a polynomial fit to free atom energies~\cite{FootnoteCoulombMatrix}.
The off-diagonal elements correspond to the Coulomb repulsion between atoms $I$ and $J$.
For a data set containing molecules with differing number of atoms, 
all the $\{\fatM\}$ of all the smaller systems are extended by zeros until they reach
the dimensionality of the largest molecule in the training set. 
The Coulomb-matrix can easily be extended to account for extended or condensed phase systems:
Let $N_J$ be the number of atoms in the unit cell, and let $N_I$ be the number of atoms in unit cell plus sufficiently large surrounding environment,
then define $M_{IJ}$ as above except that all off-diagonal elements are set to zero for all $I$ and $J$ larger than $N_J$. 

We can measure the distance between two molecules by the Euclidean norm of their diagonalized Coulomb matrices:
$d(\fatM_A,\fatM_B) = d({\bm \epsilon}_A,{\bm \epsilon}_B) = \sqrt{\sum_{I} |\epsilon_{I \in A} - \epsilon_{I \in B}|^2}$,
where ${\bm \epsilon}$ are the eigenvalues of $\fatM$ in order of decreasing absolute value.
The physical meaning of representing CCS in this way can easily be understood by considering the simplest of all molecules,
homo-nuclear diatomics (i.e.~$Z = Z_1 = Z_2$ and $r = |\fatR_1 - \fatR_2|$).
Any corresponding $\fatM$ is then simply defined by its two eigenvalues, the roots of its characteristic polynomial,
$\epsilon_{1/2} = 0.5 Z^{2.4} \pm Z^2/r$.
When measuring similarity between two such diatomics with different interatomic distances, $r_A$ and $r_B$,
the measure of similarity reduces to $d(\fatM_A,\fatM_B) = \sqrt{2}\,Z^2\,(r_B - r_A)/(r_Ar_B)$;
and the corresponding estimated potential energy curve for any new interatomic distance, $r_A$, as trained on $N$ other
interatomic distances, $\{r_i\}$, is given by
\begin{eqnarray}
E^{est}(r_A) &  = & \sum_{i=1}^N \alpha_i \exp\left[-\frac{Z^4 (r_i-r_A)^2}{\sigma^2 r^2_A r^2_i} \right].
\end{eqnarray}
In complete analogy, a ML model of the homo-nuclear dimer can also analytically be understood in terms 
of other homo-nuclear dimers with differing atomic numbers, hetero-nuclear dimers, or hetero-nuclear trimers.
The ease of differentiation with respect to not only geometry ($\partial_{r}E^{est}$) but also with respect to composition
($\partial_ZE^{est}$) illustrates further advantages of such a simple model. 

\begin{figure}
\includegraphics[scale=0.4, angle=0]{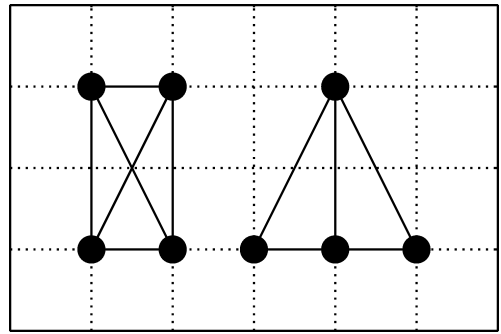}
\caption{
\label{fig:homometric}
Sketch of two homometric molecules (same stoichiometry, same sum of interatomic distances) from Ref.~\cite{TruhlarHomometric}.
The Coulomb-matrix (sorted or a set of its permutants) can distinguish these two 
molecules~\cite{NatureHomometric,MoussaComment,MoussaReply}.
}
\end{figure}

The Coulomb matrix uniquely encodes any compound because stoichiometry as well as atomic configuration are explicitly accounted for. 
Even homometric molecules~\cite{NatureHomometric}, see Fig.~\ref{fig:homometric}, are uniquely encoded by $\fatM$.
Symmetrically equivalent atoms will contribute equally, and the representation is rotationally and translationally invariant.
In order to gain invariance of $\fatM$ with respect to the index ordering of atoms one can either diagonalize, 
sort rows and columns according to their norm, or use sets of matrices with permutated rows and columns. 
Using the eigenvalues of $\fatM$ will yield an undercomplete representation. As with any coarsened representation, 
the $N_I$ degrees of freedom represented by eigenvalues will fail to uniquely represent the 
full set of $3N_I-6$ degrees of freedom for any non-linear molecule with more than three atoms~\cite{MoussaComment,MoussaReply}. 
While sorting by the norm of rows (or columns) leads to an overcomplete, index invariant, and unique representation,
the matrix is no longer differentiable for any combination of matrix entries that could be 
achieved through changes in geometry or in nuclear charges. 
Extending the representation by randomly permutated variants of Coulomb matrices is feasible, 
and leads to dramatic improvement in predictive accuracy~\cite{Montavon_NIPS2012,Montavon2012}.
To encode known invariances through such data extension has also been successful
for improving the accuracy of handwritten digit recognition~\cite{ciresan}.
Due to disadvantageous scaling, this approach might prove problematic, however, when it comes to larger systems.
As discussed in Ref.~\cite{Neuralnetworks_Behler2011}, these are all crucial criteria
for representing atomistic systems within statistical models.

\subsection{Alternative descriptors for CCS}
We shall now discuss more sophisticated alternatives to the Coulomb-matrix. 
An intuitive extension is to assume a matrix with an interatomic potential form.
This could be worth-while as long as the incurred computational overhead is small by comparison 
to the method used to generate the reference data. 
For example,
\bea
M_{IJ}^{LJ}  =
\begin{cases}
  0 & \forall \;\; I = J,\\
  \epsilon_{IJ} \left(\left(\frac{r^{eq}_{IJ}}{r_{IJ}}\right)^{12} - 2 \left(\frac{r^{eq}_{IJ}}{r_{IJ}}\right)^6\right)
  & \forall \;\; I \ne J.
\end{cases}
\label{eq:LJmatrix}
\eea
would correspond to the Lennard-Jones analog to the Coulomb-matrix.  
Similarly, a Morse or Buckingham matrix could be constructed.
One could even conceive to go beyond such pair-wise approaches and introduce interatomic 3 and higher order terms
in the form of molecular tensors.
But also electronic structure models can be encoded in terms of such a representation, such as extended H\"uckel theory, semi-empirical
quantum chemistry or tight-binding models.
For example, an orbital free Thomas-Fermi DFT representation~\cite{Review_orbitalfreeDFT2008} is possible when based on
a data-base of frozen free atomic electron densities, $\{n_I(\fatr)\}$.
The ``Hartree'' matrix is given by,
\bea
M_{IJ}^{H}  =
\begin{cases}
  0 & \forall \;\; I = J,\\
  \int d\fatr \; d\fatr' \; \frac{n_I(\fatr)n_J(\fatr')}{r_{IJ}}
  & \forall \;\; I \ne J,
\end{cases}
\label{eq:Hmatrix}
\eea
the ``external'' potential matrix is given by,
\bea
M_{IJ}^{ext}  =
\begin{cases}
  0 & \forall \;\; I = J,\\
  \int d\fatr \;  \frac{n_I(\fatr)Z_J}{r_{IJ}}
  & \forall \;\; I \ne J,
\end{cases}
\label{eq:nmatrix}
\eea
and the ``kinetic'' matrix is given within
\bea
M_{IJ}^{kin}  =
\begin{cases}
   C_F \int d\fatr\; n_I(\fatr)^{5/3}
   & \forall \;\; I = J,\\
 0  & \forall \;\; I \ne J,
\end{cases}
\label{eq:kmatrix}
\eea
where $C_F$ is a constant, and atomic integrals are evaluated over all of space. 
If need be, the kinetic matrix could even be extended by the von Weisz\"acker correction
term, $\frac{1}{8}\int d\fatr \; |\nabla n_I(\fatr)|^2/n_I(\fatr)$~\cite{Review_orbitalfreeDFT2008}.
Summation of all entries in the matrix and addition of the off-diagonal Coulomb-matrix entries would yield 
the corresponding exact DFT energy for frozen atomic electron densities.
Unfortunately, preliminary training on atomization energies of the GDB-7 data set~\cite{ReymondChemicalUniverse3} 
indicates that neither use of the Lennard-Jones nor of the Thomas Fermi matrices 
leads to any significant improvement in predictive accuracy when compared to the original
Coulomb-matrix representation in Ref.~\cite{RuppPRL2012}.
A possible explanation for this non-intuitive result is that these more sophisticated descriptors
are no longer monotonic functions in geometries and stoichiometries---in contrast to the Coulomb matrix.

An alternative new descriptor, entirely consistent with the first principles view on CCS, 
has recently been proposed~\cite{FourierDescriptor_2012}.
Each atom $I$ in the molecule is represented by its nuclear charge multiplied with a cosine term that contains
a radial distribution function of atom $I$ with respect to all other atoms $J$. 
Summing up the atomic contributions yields a Fourier series of radial distribution functions which, 
because of the superposition principle, is not only unique for each compound, 
but also invariant with respect to molecular rotations, translations, and atom indexing. 
\bea
M(d) & = & \sum_I^{N_J} Z_J^n \cos[\frac{1}{Z_J} \sum_I^{N_I} Z_I e^{-(d-d_{IJ})^2/\sigma}],
\eea
where $d_{IJ} = |\fatR_I - \fatR_J|$, and $n$ and $\sigma$ are hyper parameters that can be optimized.
This descriptor has units of charge$^n$, $d$ has units of distance and goes from zero beyond the largest interatomic distance.
As in the case of the Coulomb-matrix described above, the environment of large or condensed systems 
can be accounted for by chosing $N_I$ to be larger than $N_J$.
The reader is referred to the original paper for further details~\cite{FourierDescriptor_2012}.

\section{Concluding remarks}
We have reviewed a notion of chemical compound space that is consistent with 
any {\em ab initio} approach to atomistic simulations. 
Starting from an energy hierarchy, variations in nuclear charge distributions have been discussed,
followed by order-parameter based interpolation approaches, and statistical learning methods.
The concepts presented offer a seamless and rigorous framework to unify electronic structure theory
with rigorous rational as well as combinatorial compound design efforts.
This view of chemical space is advantageous for several reasons, 
(i) equipped with such a notion, important fundamental questions can be tackled in the future, 
including rigorous definitions of diversity in CCS, property transferability, 
uncertainty, and selection bias in training sets;
(ii) transferability and applicability typical for the black-box characteristics and the accuracy of 
{\em ab initio} calculations can be achieved;
(iii) a mathematically, physically, and chemically rigorous notion of relevant input variables
enables the application of sophisticated property optimization algorithms.
Ultimately, efforts along these lines promise to lead to ``the right compound for the right reason'', 
promising to replace by systematic engineering protocols
the heuristics and serendipity on which most, if not all, of the past compound discoveries have relied. 

\section{Acknowledgments}
The author is thankful for helpful discussions with 
C.~Anderson,
M.~Cuendet,
R.~A.~DiStasio, Jr.,
J.~R.~Hammond,
F.~Kiraly,
A.~Knoll,
G.~Montavon,
J.~E.~Moussa,
K.~R.~M\"uller,
B.~C.~Rinderspacher,
M.~Rupp,
A.~Tkatchenko,
D.~Truhlar,
M.~Tuckerman, 
A.~Vazquez-Mayagoitia.
The many participants of the 2011-program ``Navigating Chemical Compound 
Space for Materials and Bio Design'' at the Institute for Pure and Applied Mathematics, UCLA,
are also greatly acknowledged.
This research used resources of the Argonne Leadership Computing Facility at Argonne National Laboratory,
which is supported by the Office of Science of the U.S.~DOE under contract DE-AC02-06CH11357.
\bibliography{literatur}

\begin{thebibliography}{137}
\expandafter\ifx\csname natexlab\endcsname\relax\def\natexlab#1{#1}\fi
\expandafter\ifx\csname bibnamefont\endcsname\relax
  \def\bibnamefont#1{#1}\fi
\expandafter\ifx\csname bibfnamefont\endcsname\relax
  \def\bibfnamefont#1{#1}\fi
\expandafter\ifx\csname citenamefont\endcsname\relax
  \def\citenamefont#1{#1}\fi
\expandafter\ifx\csname url\endcsname\relax
  \def\url#1{\texttt{#1}}\fi
\expandafter\ifx\csname urlprefix\endcsname\relax\def\urlprefix{URL }\fi
\providecommand{\bibinfo}[2]{#2}
\providecommand{\eprint}[2][]{\url{#2}}

\bibitem[{\citenamefont{Kirkpatrick and Ellis}(2004)}]{ChemicalSpace}
\bibinfo{author}{\bibfnamefont{P.}~\bibnamefont{Kirkpatrick}} \bibnamefont{and}
  \bibinfo{author}{\bibfnamefont{C.}~\bibnamefont{Ellis}},
  \bibinfo{journal}{Nature} \textbf{\bibinfo{volume}{432}},
  \bibinfo{pages}{823} (\bibinfo{year}{2004}).

\bibitem[{\citenamefont{Burke}(2011)}]{AbInitioDefinitionByKieronBurke}
\bibinfo{author}{\bibfnamefont{K.}~\bibnamefont{Burke}} (\bibinfo{year}{2011}),
  \bibinfo{note}{''Any method whose parametrization does not depend on the
  chemical system being studied can be called an {\em ab initio} method.'' Oral
  communication, IPAM, UCLA.}

\bibitem[{\citenamefont{Helgaker et~al.}(2000)\citenamefont{Helgaker,
  J{\o}rgensen, and Olsen}}]{MolecularElectronicStructureTheory}
\bibinfo{author}{\bibfnamefont{T.}~\bibnamefont{Helgaker}},
  \bibinfo{author}{\bibfnamefont{P.}~\bibnamefont{J{\o}rgensen}},
  \bibnamefont{and} \bibinfo{author}{\bibfnamefont{J.}~\bibnamefont{Olsen}},
  \emph{\bibinfo{title}{Molecular Electronic-Structure Theory}}
  (\bibinfo{publisher}{John Wiley \& Sons, LTD}, \bibinfo{year}{2000}).

\bibitem[{\citenamefont{Tuckerman}(2010)}]{tuckerman_book_SM}
\bibinfo{author}{\bibfnamefont{M.~E.} \bibnamefont{Tuckerman}},
  \emph{\bibinfo{title}{Statistical mechanics: Theory and molecular
  simulation}} (\bibinfo{publisher}{Oxford University Press},
  \bibinfo{year}{2010}).

\bibitem[{\citenamefont{Hohenberg and Kohn}(1964)}]{HK}
\bibinfo{author}{\bibfnamefont{P.}~\bibnamefont{Hohenberg}} \bibnamefont{and}
  \bibinfo{author}{\bibfnamefont{W.}~\bibnamefont{Kohn}},
  \bibinfo{journal}{Phys. Rev.} \textbf{\bibinfo{volume}{136}},
  \bibinfo{pages}{B864} (\bibinfo{year}{1964}).

\bibitem[{\citenamefont{Becke}(1988)}]{B88X}
\bibinfo{author}{\bibfnamefont{A.~D.} \bibnamefont{Becke}},
  \bibinfo{journal}{Phys. Rev. A} \textbf{\bibinfo{volume}{38}},
  \bibinfo{pages}{3098} (\bibinfo{year}{1988}).

\bibitem[{\citenamefont{Lee et~al.}(1988)\citenamefont{Lee, Yang, and
  Parr}}]{lyp}
\bibinfo{author}{\bibfnamefont{C.}~\bibnamefont{Lee}},
  \bibinfo{author}{\bibfnamefont{W.}~\bibnamefont{Yang}}, \bibnamefont{and}
  \bibinfo{author}{\bibfnamefont{R.~G.} \bibnamefont{Parr}},
  \bibinfo{journal}{Phys. Rev. B} \textbf{\bibinfo{volume}{37}},
  \bibinfo{pages}{785} (\bibinfo{year}{1988}).

\bibitem[{\citenamefont{Henze and Blair}(1931)}]{EnumerateAlkene}
\bibinfo{author}{\bibfnamefont{H.~R.} \bibnamefont{Henze}} \bibnamefont{and}
  \bibinfo{author}{\bibfnamefont{C.~M.} \bibnamefont{Blair}},
  \bibinfo{journal}{J. Am. Chem. Soc.} \textbf{\bibinfo{volume}{53}},
  \bibinfo{pages}{3077} (\bibinfo{year}{1931}).

\bibitem[{\citenamefont{Perry}(1932)}]{EnumerateAlkene2}
\bibinfo{author}{\bibfnamefont{D.}~\bibnamefont{Perry}}, \bibinfo{journal}{J.
  Am. Chem. Soc.} \textbf{\bibinfo{volume}{54}}, \bibinfo{pages}{2918}
  (\bibinfo{year}{1932}).

\bibitem[{\citenamefont{Bytautas and Klein}(1998)}]{EnumerateAlkene3}
\bibinfo{author}{\bibfnamefont{L.}~\bibnamefont{Bytautas}} \bibnamefont{and}
  \bibinfo{author}{\bibfnamefont{D.}~\bibnamefont{Klein}}, \bibinfo{journal}{J.
  Chem. Inf. Comp. Sci.} \textbf{\bibinfo{volume}{38}}, \bibinfo{pages}{1063}
  (\bibinfo{year}{1998}).

\bibitem[{\citenamefont{Braun et~al.}(2004)\citenamefont{Braun, Gugisch,
  Kerber, Laue, Meringer, and R\"ucker}}]{CanonizerMeringer2004}
\bibinfo{author}{\bibfnamefont{J.}~\bibnamefont{Braun}},
  \bibinfo{author}{\bibfnamefont{R.}~\bibnamefont{Gugisch}},
  \bibinfo{author}{\bibfnamefont{A.}~\bibnamefont{Kerber}},
  \bibinfo{author}{\bibfnamefont{R.}~\bibnamefont{Laue}},
  \bibinfo{author}{\bibfnamefont{M.}~\bibnamefont{Meringer}}, \bibnamefont{and}
  \bibinfo{author}{\bibfnamefont{C.}~\bibnamefont{R\"ucker}},
  \bibinfo{journal}{J. Chem. Inf. Comp. Sci.} \textbf{\bibinfo{volume}{44}},
  \bibinfo{pages}{542} (\bibinfo{year}{2004}).

\bibitem[{\citenamefont{Blum and Reymond}(2009)}]{ReymondChemicalUniverse3}
\bibinfo{author}{\bibfnamefont{L.~C.} \bibnamefont{Blum}} \bibnamefont{and}
  \bibinfo{author}{\bibfnamefont{J.-L.} \bibnamefont{Reymond}},
  \bibinfo{journal}{J. Am. Chem. Soc.} \textbf{\bibinfo{volume}{131}},
  \bibinfo{pages}{8732} (\bibinfo{year}{2009}).

\bibitem[{\citenamefont{van Duin et~al.}(2001)\citenamefont{van Duin, Dasgupta,
  Lorant, , and {Goddard III}}}]{ReaxFF2001}
\bibinfo{author}{\bibfnamefont{A.~C.~T.} \bibnamefont{van Duin}},
  \bibinfo{author}{\bibfnamefont{S.}~\bibnamefont{Dasgupta}},
  \bibinfo{author}{\bibfnamefont{F.}~\bibnamefont{Lorant}}, , \bibnamefont{and}
  \bibinfo{author}{\bibfnamefont{W.~A.} \bibnamefont{{Goddard III}}},
  \bibinfo{journal}{J. Phys. Chem. A} \textbf{\bibinfo{volume}{105}},
  \bibinfo{pages}{9396} (\bibinfo{year}{2001}).

\bibitem[{\citenamefont{Rapp\'e et~al.}(1992)\citenamefont{Rapp\'e, Casewit,
  Colwell, {Goddard III}, and Skid}}]{UFFRappeGoddard1992}
\bibinfo{author}{\bibfnamefont{A.~K.} \bibnamefont{Rapp\'e}},
  \bibinfo{author}{\bibfnamefont{C.~J.} \bibnamefont{Casewit}},
  \bibinfo{author}{\bibfnamefont{K.~S.} \bibnamefont{Colwell}},
  \bibinfo{author}{\bibfnamefont{W.~A.} \bibnamefont{{Goddard III}}},
  \bibnamefont{and} \bibinfo{author}{\bibfnamefont{W.~M.} \bibnamefont{Skid}},
  \bibinfo{journal}{J. Am. Chem. Soc.} \textbf{\bibinfo{volume}{114}},
  \bibinfo{pages}{10024} (\bibinfo{year}{1992}).

\bibitem[{\citenamefont{Geerlings et~al.}(2003)\citenamefont{Geerlings, Proft,
  and Langenaeker}}]{Geerlings_DFTConcepts}
\bibinfo{author}{\bibfnamefont{P.}~\bibnamefont{Geerlings}},
  \bibinfo{author}{\bibfnamefont{F.~D.} \bibnamefont{Proft}}, \bibnamefont{and}
  \bibinfo{author}{\bibfnamefont{W.}~\bibnamefont{Langenaeker}},
  \bibinfo{journal}{Chem. Rev.} \textbf{\bibinfo{volume}{103}},
  \bibinfo{pages}{1793} (\bibinfo{year}{2003}).

\bibitem[{\citenamefont{Parr and Yang}(1989)}]{parryang}
\bibinfo{author}{\bibfnamefont{R.~G.} \bibnamefont{Parr}} \bibnamefont{and}
  \bibinfo{author}{\bibfnamefont{W.}~\bibnamefont{Yang}},
  \emph{\bibinfo{title}{Density functional theory of atoms and molecules}}
  (\bibinfo{publisher}{Oxford Science Publications}, \bibinfo{year}{1989}).

\bibitem[{\citenamefont{von Lilienfeld and
  Tuckerman}(2006)}]{anatole-jcp2006-2}
\bibinfo{author}{\bibfnamefont{O.~A.} \bibnamefont{von Lilienfeld}}
  \bibnamefont{and} \bibinfo{author}{\bibfnamefont{M.~E.}
  \bibnamefont{Tuckerman}}, \bibinfo{journal}{J. Chem. Phys.}
  \textbf{\bibinfo{volume}{125}}, \bibinfo{pages}{154104}
  (\bibinfo{year}{2006}).

\bibitem[{\citenamefont{Capitani et~al.}(1982)\citenamefont{Capitani,
  Nalewajski, and Parr}}]{nonBO_DFT}
\bibinfo{author}{\bibfnamefont{J.~F.} \bibnamefont{Capitani}},
  \bibinfo{author}{\bibfnamefont{R.~F.} \bibnamefont{Nalewajski}},
  \bibnamefont{and} \bibinfo{author}{\bibfnamefont{R.~G.} \bibnamefont{Parr}},
  \bibinfo{journal}{J. Chem. Phys.} \textbf{\bibinfo{volume}{76}},
  \bibinfo{pages}{568} (\bibinfo{year}{1982}).

\bibitem[{\citenamefont{von Lilienfeld
  et~al.}(2005{\natexlab{a}})\citenamefont{von Lilienfeld, Lins, and
  Rothlisberger}}]{anatole-prl2005}
\bibinfo{author}{\bibfnamefont{O.~A.} \bibnamefont{von Lilienfeld}},
  \bibinfo{author}{\bibfnamefont{R.}~\bibnamefont{Lins}}, \bibnamefont{and}
  \bibinfo{author}{\bibfnamefont{U.}~\bibnamefont{Rothlisberger}},
  \bibinfo{journal}{Phys. Rev. Lett.} \textbf{\bibinfo{volume}{95}},
  \bibinfo{pages}{153002} (\bibinfo{year}{2005}{\natexlab{a}}).

\bibitem[{\citenamefont{Marcon et~al.}(2007)\citenamefont{Marcon, von
  Lilienfeld, and Andrienko}}]{anatole-jcp2007}
\bibinfo{author}{\bibfnamefont{V.}~\bibnamefont{Marcon}},
  \bibinfo{author}{\bibfnamefont{O.~A.} \bibnamefont{von Lilienfeld}},
  \bibnamefont{and}
  \bibinfo{author}{\bibfnamefont{D.}~\bibnamefont{Andrienko}},
  \bibinfo{journal}{J. Chem. Phys.} \textbf{\bibinfo{volume}{127}},
  \bibinfo{pages}{064305} (\bibinfo{year}{2007}).

\bibitem[{\citenamefont{Kirkwood}(1935)}]{TI}
\bibinfo{author}{\bibfnamefont{J.~G.} \bibnamefont{Kirkwood}},
  \bibinfo{journal}{J. Chem. Phys.} \textbf{\bibinfo{volume}{3}},
  \bibinfo{pages}{300} (\bibinfo{year}{1935}).

\bibitem[{\citenamefont{von Lilienfeld and Tuckerman}(2007)}]{anatole-jctc2007}
\bibinfo{author}{\bibfnamefont{O.~A.} \bibnamefont{von Lilienfeld}}
  \bibnamefont{and} \bibinfo{author}{\bibfnamefont{M.~E.}
  \bibnamefont{Tuckerman}}, \bibinfo{journal}{J. Chem. Theory Comput.}
  \textbf{\bibinfo{volume}{3}}, \bibinfo{pages}{1083} (\bibinfo{year}{2007}).

\bibitem[{\citenamefont{Hellmann}(1935)}]{Hellmann1}
\bibinfo{author}{\bibfnamefont{H.}~\bibnamefont{Hellmann}},
  \bibinfo{journal}{J. Chem. Phys.} \textbf{\bibinfo{volume}{3}},
  \bibinfo{pages}{61} (\bibinfo{year}{1935}).

\bibitem[{\citenamefont{Hellmann}(1936)}]{Hellmann2}
\bibinfo{author}{\bibfnamefont{H.}~\bibnamefont{Hellmann}},
  \bibinfo{journal}{J. Chem. Phys.} \textbf{\bibinfo{volume}{4}},
  \bibinfo{pages}{324} (\bibinfo{year}{1936}).

\bibitem[{\citenamefont{Phillips and Kleinman}(1959)}]{KleinmansPP}
\bibinfo{author}{\bibfnamefont{J.~C.} \bibnamefont{Phillips}} \bibnamefont{and}
  \bibinfo{author}{\bibfnamefont{L.}~\bibnamefont{Kleinman}},
  \bibinfo{journal}{Phys. Rev.} \textbf{\bibinfo{volume}{116}},
  \bibinfo{pages}{287} (\bibinfo{year}{1959}).

\bibitem[{\citenamefont{Weeks and Rice}(1968)}]{WeeksPP}
\bibinfo{author}{\bibfnamefont{J.~D.} \bibnamefont{Weeks}} \bibnamefont{and}
  \bibinfo{author}{\bibfnamefont{S.~A.} \bibnamefont{Rice}},
  \bibinfo{journal}{J. Chem. Phys.} \textbf{\bibinfo{volume}{49}},
  \bibinfo{pages}{2741} (\bibinfo{year}{1968}).

\bibitem[{\citenamefont{Bachelet et~al.}(1982)\citenamefont{Bachelet, Hamann,
  and Schluter}}]{bhs-with-title}
\bibinfo{author}{\bibfnamefont{G.~B.} \bibnamefont{Bachelet}},
  \bibinfo{author}{\bibfnamefont{D.~R.} \bibnamefont{Hamann}},
  \bibnamefont{and} \bibinfo{author}{\bibfnamefont{M.}~\bibnamefont{Schluter}},
  \bibinfo{journal}{Phys. Rev. B} \textbf{\bibinfo{volume}{26}},
  \bibinfo{pages}{4199} (\bibinfo{year}{1982}).

\bibitem[{\citenamefont{Christiansen et~al.}(1979)\citenamefont{Christiansen,
  Lee, and Pitzer}}]{ChristiansensPP}
\bibinfo{author}{\bibfnamefont{P.~A.} \bibnamefont{Christiansen}},
  \bibinfo{author}{\bibfnamefont{Y.~S.} \bibnamefont{Lee}}, \bibnamefont{and}
  \bibinfo{author}{\bibfnamefont{K.~S.} \bibnamefont{Pitzer}},
  \bibinfo{journal}{J. Chem. Phys.} \textbf{\bibinfo{volume}{71}},
  \bibinfo{pages}{4445} (\bibinfo{year}{1979}).

\bibitem[{\citenamefont{Pulay}(1969)}]{Pulay-force}
\bibinfo{author}{\bibfnamefont{P.}~\bibnamefont{Pulay}}, \bibinfo{journal}{Mol.
  Phys.} \textbf{\bibinfo{volume}{229}} (\bibinfo{year}{1969}).

\bibitem[{\citenamefont{Hartwigsen et~al.}(1998)\citenamefont{Hartwigsen,
  Goedecker, and Hutter}}]{sgpsp}
\bibinfo{author}{\bibfnamefont{C.}~\bibnamefont{Hartwigsen}},
  \bibinfo{author}{\bibfnamefont{S.}~\bibnamefont{Goedecker}},
  \bibnamefont{and} \bibinfo{author}{\bibfnamefont{J.}~\bibnamefont{Hutter}},
  \bibinfo{journal}{Phys. Rev. B} \textbf{\bibinfo{volume}{58}},
  \bibinfo{pages}{3641} (\bibinfo{year}{1998}).

\bibitem[{\citenamefont{Rieger and Vogl}(1995)}]{SICPP-Vogl}
\bibinfo{author}{\bibfnamefont{M.~M.} \bibnamefont{Rieger}} \bibnamefont{and}
  \bibinfo{author}{\bibfnamefont{P.}~\bibnamefont{Vogl}},
  \bibinfo{journal}{Phys. Rev. B} \textbf{\bibinfo{volume}{52}},
  \bibinfo{pages}{16567} (\bibinfo{year}{1995}).

\bibitem[{\citenamefont{Baumeier et~al.}(2006)\citenamefont{Baumeier, Kr\"uger,
  and Pollmann}}]{SICPP-Pollmann}
\bibinfo{author}{\bibfnamefont{B.}~\bibnamefont{Baumeier}},
  \bibinfo{author}{\bibfnamefont{P.}~\bibnamefont{Kr\"uger}}, \bibnamefont{and}
  \bibinfo{author}{\bibfnamefont{J.}~\bibnamefont{Pollmann}},
  \bibinfo{journal}{Phys. Rev. B} \textbf{\bibinfo{volume}{73}},
  \bibinfo{pages}{195205} (\bibinfo{year}{2006}).

\bibitem[{\citenamefont{von Lilienfeld
  et~al.}(2005{\natexlab{b}})\citenamefont{von Lilienfeld, Tavernelli,
  Rothlisberger, and Sebastiani}}]{anatole-jcp2005}
\bibinfo{author}{\bibfnamefont{O.~A.} \bibnamefont{von Lilienfeld}},
  \bibinfo{author}{\bibfnamefont{I.}~\bibnamefont{Tavernelli}},
  \bibinfo{author}{\bibfnamefont{U.}~\bibnamefont{Rothlisberger}},
  \bibnamefont{and}
  \bibinfo{author}{\bibfnamefont{D.}~\bibnamefont{Sebastiani}},
  \bibinfo{journal}{J. Chem. Phys.} \textbf{\bibinfo{volume}{122}},
  \bibinfo{pages}{014113} (\bibinfo{year}{2005}{\natexlab{b}}).

\bibitem[{\citenamefont{DiLabio et~al.}(2002)\citenamefont{DiLabio, Hurley, and
  Christiansen}}]{Christiansen-JCP-2002}
\bibinfo{author}{\bibfnamefont{G.~A.} \bibnamefont{DiLabio}},
  \bibinfo{author}{\bibfnamefont{M.~M.} \bibnamefont{Hurley}},
  \bibnamefont{and} \bibinfo{author}{\bibfnamefont{P.~A.}
  \bibnamefont{Christiansen}}, \bibinfo{journal}{J. Chem. Phys.}
  \textbf{\bibinfo{volume}{116}}, \bibinfo{pages}{9578} (\bibinfo{year}{2002}).

\bibitem[{\citenamefont{von Lilienfeld et~al.}(2004)\citenamefont{von
  Lilienfeld, Tavernelli, Rothlisberger, and Sebastiani}}]{anatole-prl2004}
\bibinfo{author}{\bibfnamefont{O.~A.} \bibnamefont{von Lilienfeld}},
  \bibinfo{author}{\bibfnamefont{I.}~\bibnamefont{Tavernelli}},
  \bibinfo{author}{\bibfnamefont{U.}~\bibnamefont{Rothlisberger}},
  \bibnamefont{and}
  \bibinfo{author}{\bibfnamefont{D.}~\bibnamefont{Sebastiani}},
  \bibinfo{journal}{Phys. Rev. Lett.} \textbf{\bibinfo{volume}{93}},
  \bibinfo{pages}{153004} (\bibinfo{year}{2004}).

\bibitem[{\citenamefont{Torres and DiLabio}(2012)}]{DiLabioDCACP2012}
\bibinfo{author}{\bibfnamefont{E.}~\bibnamefont{Torres}} \bibnamefont{and}
  \bibinfo{author}{\bibfnamefont{G.~A.} \bibnamefont{DiLabio}},
  \bibinfo{journal}{J. Phys. Chem. Lett.} \textbf{\bibinfo{volume}{3}},
  \bibinfo{pages}{1738} (\bibinfo{year}{2012}).

\bibitem[{\citenamefont{Christensen}(1984)}]{Christensen1984}
\bibinfo{author}{\bibfnamefont{N.~E.} \bibnamefont{Christensen}},
  \bibinfo{journal}{Phys. Rev. B} \textbf{\bibinfo{volume}{30}},
  \bibinfo{pages}{5753} (\bibinfo{year}{1984}).

\bibitem[{\citenamefont{{D. Segev and A. Janotti and C. G. Van de
  Walle}}(2007)}]{vandewalle2007}
\bibinfo{author}{\bibnamefont{{D. Segev and A. Janotti and C. G. Van de
  Walle}}}, \bibinfo{journal}{Phys. Rev. B} \textbf{\bibinfo{volume}{75}},
  \bibinfo{pages}{35201} (\bibinfo{year}{2007}).

\bibitem[{\citenamefont{Leung et~al.}(2009)\citenamefont{Leung, Rempe, and von
  Lilienfeld}}]{anatole-jcp2009}
\bibinfo{author}{\bibfnamefont{K.}~\bibnamefont{Leung}},
  \bibinfo{author}{\bibfnamefont{S.~B.} \bibnamefont{Rempe}}, \bibnamefont{and}
  \bibinfo{author}{\bibfnamefont{O.~A.} \bibnamefont{von Lilienfeld}},
  \bibinfo{journal}{J. Chem. Phys.} \textbf{\bibinfo{volume}{130}},
  \bibinfo{pages}{204507} (\bibinfo{year}{2009}).

\bibitem[{\citenamefont{Sulpizi and Sprik}(2008)}]{pKaLore2008}
\bibinfo{author}{\bibfnamefont{M.}~\bibnamefont{Sulpizi}} \bibnamefont{and}
  \bibinfo{author}{\bibfnamefont{M.}~\bibnamefont{Sprik}},
  \bibinfo{journal}{Phys. Chem. Chem. Phys.} \textbf{\bibinfo{volume}{10}},
  \bibinfo{pages}{5238} (\bibinfo{year}{2008}).

\bibitem[{\citenamefont{Alf\`e et~al.}(2000)\citenamefont{Alf\`e, Gillan, and
  Price}}]{Alfe-nature2000}
\bibinfo{author}{\bibfnamefont{D.}~\bibnamefont{Alf\`e}},
  \bibinfo{author}{\bibfnamefont{M.~J.} \bibnamefont{Gillan}},
  \bibnamefont{and} \bibinfo{author}{\bibfnamefont{G.~D.} \bibnamefont{Price}},
  \bibinfo{journal}{Nature} \textbf{\bibinfo{volume}{405}},
  \bibinfo{pages}{172} (\bibinfo{year}{2000}).

\bibitem[{\citenamefont{Sheppard et~al.}(2010)\citenamefont{Sheppard,
  Henkelman, and von Lilienfeld}}]{CatalystSheppard2010}
\bibinfo{author}{\bibfnamefont{D.}~\bibnamefont{Sheppard}},
  \bibinfo{author}{\bibfnamefont{G.}~\bibnamefont{Henkelman}},
  \bibnamefont{and} \bibinfo{author}{\bibfnamefont{O.~A.} \bibnamefont{von
  Lilienfeld}}, \bibinfo{journal}{J. Chem. Phys.}
  \textbf{\bibinfo{volume}{133}}, \bibinfo{pages}{084104}
  (\bibinfo{year}{2010}).

\bibitem[{\citenamefont{Goedecker et~al.}(1996)\citenamefont{Goedecker, Teter,
  and Hutter}}]{SG}
\bibinfo{author}{\bibfnamefont{S.}~\bibnamefont{Goedecker}},
  \bibinfo{author}{\bibfnamefont{M.}~\bibnamefont{Teter}}, \bibnamefont{and}
  \bibinfo{author}{\bibfnamefont{J.}~\bibnamefont{Hutter}},
  \bibinfo{journal}{Phys. Rev. B} \textbf{\bibinfo{volume}{54}},
  \bibinfo{pages}{1703} (\bibinfo{year}{1996}).

\bibitem[{\citenamefont{Krack}(2005)}]{KrackPP}
\bibinfo{author}{\bibfnamefont{M.}~\bibnamefont{Krack}},
  \bibinfo{journal}{Theor. Chim. Acta} \textbf{\bibinfo{volume}{114}},
  \bibinfo{pages}{145} (\bibinfo{year}{2005}).

\bibitem[{\citenamefont{Weigend et~al.}(2004)\citenamefont{Weigend, Schrodt,
  and
  Ahlrichs}}]{AlchemicalDerivativeBinaryMetalCluster_WeigendSchrodtAhlrichs200%
4}
\bibinfo{author}{\bibfnamefont{F.}~\bibnamefont{Weigend}},
  \bibinfo{author}{\bibfnamefont{C.}~\bibnamefont{Schrodt}}, \bibnamefont{and}
  \bibinfo{author}{\bibfnamefont{R.}~\bibnamefont{Ahlrichs}},
  \bibinfo{journal}{J. Chem. Phys.} \textbf{\bibinfo{volume}{121}},
  \bibinfo{pages}{10380} (\bibinfo{year}{2004}).

\bibitem[{\citenamefont{Cardenas et~al.}(2011)\citenamefont{Cardenas, Tiznado,
  Ayers, and Fuentealba}}]{CardenasFukui2011}
\bibinfo{author}{\bibfnamefont{C.}~\bibnamefont{Cardenas}},
  \bibinfo{author}{\bibfnamefont{W.}~\bibnamefont{Tiznado}},
  \bibinfo{author}{\bibfnamefont{P.~W.} \bibnamefont{Ayers}}, \bibnamefont{and}
  \bibinfo{author}{\bibfnamefont{P.}~\bibnamefont{Fuentealba}},
  \bibinfo{journal}{J. Phys. Chem. A} \textbf{\bibinfo{volume}{115}},
  \bibinfo{pages}{2325} (\bibinfo{year}{2011}).

\bibitem[{\citenamefont{Lesiuk et~al.}(2012)\citenamefont{Lesiuk, Balawender,
  and Zachara}}]{LesiukHigherOrderAlchemy2012}
\bibinfo{author}{\bibfnamefont{M.}~\bibnamefont{Lesiuk}},
  \bibinfo{author}{\bibfnamefont{R.}~\bibnamefont{Balawender}},
  \bibnamefont{and} \bibinfo{author}{\bibfnamefont{J.}~\bibnamefont{Zachara}},
  \bibinfo{journal}{J. Chem. Phys.} \textbf{\bibinfo{volume}{136}},
  \bibinfo{pages}{034104} (\bibinfo{year}{2012}).

\bibitem[{\citenamefont{Wang et~al.}(2006)\citenamefont{Wang, Hu, Beratan, and
  Yang}}]{RCD_Yang2006}
\bibinfo{author}{\bibfnamefont{M.}~\bibnamefont{Wang}},
  \bibinfo{author}{\bibfnamefont{X.}~\bibnamefont{Hu}},
  \bibinfo{author}{\bibfnamefont{D.~N.} \bibnamefont{Beratan}},
  \bibnamefont{and} \bibinfo{author}{\bibfnamefont{W.}~\bibnamefont{Yang}},
  \bibinfo{journal}{J. Am. Chem. Soc.} \textbf{\bibinfo{volume}{128}},
  \bibinfo{pages}{3228} (\bibinfo{year}{2006}).

\bibitem[{\citenamefont{Xiao et~al.}(2008)\citenamefont{Xiao, Yang, and
  Beratan}}]{InverseBeratanYang2008}
\bibinfo{author}{\bibfnamefont{D.}~\bibnamefont{Xiao}},
  \bibinfo{author}{\bibfnamefont{W.}~\bibnamefont{Yang}}, \bibnamefont{and}
  \bibinfo{author}{\bibfnamefont{D.~N.} \bibnamefont{Beratan}},
  \bibinfo{journal}{J. Chem. Phys.} \textbf{\bibinfo{volume}{129}},
  \bibinfo{pages}{044106} (\bibinfo{year}{2008}).

\bibitem[{\citenamefont{Hu et~al.}(2008)\citenamefont{Hu, Beratan, and
  Yang}}]{GradientMCBeratanYang2008}
\bibinfo{author}{\bibfnamefont{X.}~\bibnamefont{Hu}},
  \bibinfo{author}{\bibfnamefont{D.~N.} \bibnamefont{Beratan}},
  \bibnamefont{and} \bibinfo{author}{\bibfnamefont{W.}~\bibnamefont{Yang}},
  \bibinfo{journal}{J. Chem. Phys.} \textbf{\bibinfo{volume}{129}},
  \bibinfo{pages}{064102} (\bibinfo{year}{2008}).

\bibitem[{\citenamefont{Balamurugan et~al.}(2008)\citenamefont{Balamurugan,
  Yang, and Beratan}}]{HybridExplorationCCSBeratanYang2008}
\bibinfo{author}{\bibfnamefont{D.}~\bibnamefont{Balamurugan}},
  \bibinfo{author}{\bibfnamefont{W.}~\bibnamefont{Yang}}, \bibnamefont{and}
  \bibinfo{author}{\bibfnamefont{D.~N.} \bibnamefont{Beratan}},
  \bibinfo{journal}{J. Chem. Phys.} \textbf{\bibinfo{volume}{129}},
  \bibinfo{pages}{174105} (\bibinfo{year}{2008}).

\bibitem[{\citenamefont{Keinan et~al.}(2008)\citenamefont{Keinan, Therien,
  Beratan, and Yang}}]{DesignKeinanBeratanYang2008}
\bibinfo{author}{\bibfnamefont{S.}~\bibnamefont{Keinan}},
  \bibinfo{author}{\bibfnamefont{M.~J.} \bibnamefont{Therien}},
  \bibinfo{author}{\bibfnamefont{D.~N.} \bibnamefont{Beratan}},
  \bibnamefont{and} \bibinfo{author}{\bibfnamefont{W.}~\bibnamefont{Yang}},
  \bibinfo{journal}{J. Phys. Chem. A} \textbf{\bibinfo{volume}{112}},
  \bibinfo{pages}{12203} (\bibinfo{year}{2008}).

\bibitem[{\citenamefont{Rinderspacher et~al.}(2009)\citenamefont{Rinderspacher,
  Andzelm, Rawlett, Dougherty, Beratan, and
  Yang}}]{MolecularDesignRinderspacherBeratanYang2009}
\bibinfo{author}{\bibfnamefont{B.~C.} \bibnamefont{Rinderspacher}},
  \bibinfo{author}{\bibfnamefont{J.}~\bibnamefont{Andzelm}},
  \bibinfo{author}{\bibfnamefont{A.}~\bibnamefont{Rawlett}},
  \bibinfo{author}{\bibfnamefont{J.}~\bibnamefont{Dougherty}},
  \bibinfo{author}{\bibfnamefont{D.~N.} \bibnamefont{Beratan}},
  \bibnamefont{and} \bibinfo{author}{\bibfnamefont{W.}~\bibnamefont{Yang}},
  \bibinfo{journal}{J. Chem. Theory Comput.} \textbf{\bibinfo{volume}{5}},
  \bibinfo{pages}{3321} (\bibinfo{year}{2009}).

\bibitem[{\citenamefont{Marder et~al.}(1991)\citenamefont{Marder, Beratan, and
  Cheng}}]{Beratan1991}
\bibinfo{author}{\bibfnamefont{S.~R.} \bibnamefont{Marder}},
  \bibinfo{author}{\bibfnamefont{D.~N.} \bibnamefont{Beratan}},
  \bibnamefont{and} \bibinfo{author}{\bibfnamefont{L.-T.} \bibnamefont{Cheng}},
  \bibinfo{journal}{Science} \textbf{\bibinfo{volume}{252}},
  \bibinfo{pages}{103} (\bibinfo{year}{1991}).

\bibitem[{\citenamefont{Kuhn and Beratan}(1996)}]{Beratan1996}
\bibinfo{author}{\bibfnamefont{C.}~\bibnamefont{Kuhn}} \bibnamefont{and}
  \bibinfo{author}{\bibfnamefont{D.~N.} \bibnamefont{Beratan}},
  \bibinfo{journal}{J. Phys. Chem.} \textbf{\bibinfo{volume}{100}},
  \bibinfo{pages}{10595} (\bibinfo{year}{1996}).

\bibitem[{\citenamefont{d'Avezac and Zunger}(2008)}]{AvezacZunger-prb2008}
\bibinfo{author}{\bibfnamefont{M.}~\bibnamefont{d'Avezac}} \bibnamefont{and}
  \bibinfo{author}{\bibfnamefont{A.}~\bibnamefont{Zunger}},
  \bibinfo{journal}{Phys. Rev. B} \textbf{\bibinfo{volume}{78}},
  \bibinfo{pages}{064102} (\bibinfo{year}{2008}).

\bibitem[{\citenamefont{Sablon et~al.}(2010)\citenamefont{Sablon, Proft, Ayers,
  and Geerlings}}]{2ndOrderDerivsWrtVext_DeProftAyersGeerlings}
\bibinfo{author}{\bibfnamefont{N.}~\bibnamefont{Sablon}},
  \bibinfo{author}{\bibfnamefont{F.~D.} \bibnamefont{Proft}},
  \bibinfo{author}{\bibfnamefont{P.~W.} \bibnamefont{Ayers}}, \bibnamefont{and}
  \bibinfo{author}{\bibfnamefont{P.}~\bibnamefont{Geerlings}},
  \bibinfo{journal}{J. Chem. Theory Comput.} \textbf{\bibinfo{volume}{6}},
  \bibinfo{pages}{3671} (\bibinfo{year}{2010}).

\bibitem[{\citenamefont{Yang et~al.}(2012)\citenamefont{Yang, Cohen, Proft, and
  Geerlings}}]{YangCohenGeerlingsAnalyticalFukui2012}
\bibinfo{author}{\bibfnamefont{W.}~\bibnamefont{Yang}},
  \bibinfo{author}{\bibfnamefont{A.~J.} \bibnamefont{Cohen}},
  \bibinfo{author}{\bibfnamefont{F.~D.} \bibnamefont{Proft}}, \bibnamefont{and}
  \bibinfo{author}{\bibfnamefont{P.}~\bibnamefont{Geerlings}},
  \bibinfo{journal}{J. Chem. Phys.} \textbf{\bibinfo{volume}{136}},
  \bibinfo{pages}{144110} (\bibinfo{year}{2012}).

\bibitem[{\citenamefont{Perdew et~al.}(1982)\citenamefont{Perdew, Parr, Levy,
  and Balduz}}]{DD_Perdew1}
\bibinfo{author}{\bibfnamefont{J.~P.} \bibnamefont{Perdew}},
  \bibinfo{author}{\bibfnamefont{R.~G.} \bibnamefont{Parr}},
  \bibinfo{author}{\bibfnamefont{M.}~\bibnamefont{Levy}}, \bibnamefont{and}
  \bibinfo{author}{\bibfnamefont{J.~L.} \bibnamefont{Balduz}},
  \bibinfo{journal}{Phys. Rev. Lett.} \textbf{\bibinfo{volume}{49}},
  \bibinfo{pages}{1691} (\bibinfo{year}{1982}).

\bibitem[{\citenamefont{Perdew and Levy}(1983)}]{DD_Perdew2}
\bibinfo{author}{\bibfnamefont{J.~P.} \bibnamefont{Perdew}} \bibnamefont{and}
  \bibinfo{author}{\bibfnamefont{M.}~\bibnamefont{Levy}},
  \bibinfo{journal}{Phys. Rev. Lett.} \textbf{\bibinfo{volume}{51}},
  \bibinfo{pages}{1884} (\bibinfo{year}{1983}).

\bibitem[{\citenamefont{Mori-S\'anchez
  et~al.}(2009)\citenamefont{Mori-S\'anchez, Cohen, and
  Yang}}]{DiscontinuousXC_MoriSanchezCohenYang2009}
\bibinfo{author}{\bibfnamefont{P.}~\bibnamefont{Mori-S\'anchez}},
  \bibinfo{author}{\bibfnamefont{A.~J.} \bibnamefont{Cohen}}, \bibnamefont{and}
  \bibinfo{author}{\bibfnamefont{W.}~\bibnamefont{Yang}},
  \bibinfo{journal}{Phys. Rev. Lett.} \textbf{\bibinfo{volume}{102}},
  \bibinfo{pages}{066403} (\bibinfo{year}{2009}).

\bibitem[{\citenamefont{Constantin et~al.}(2010)\citenamefont{Constantin,
  Snyder, Perdew, and Burke}}]{HighZ_Burke2010}
\bibinfo{author}{\bibfnamefont{L.~A.} \bibnamefont{Constantin}},
  \bibinfo{author}{\bibfnamefont{J.~C.} \bibnamefont{Snyder}},
  \bibinfo{author}{\bibfnamefont{J.~P.} \bibnamefont{Perdew}},
  \bibnamefont{and} \bibinfo{author}{\bibfnamefont{K.}~\bibnamefont{Burke}},
  \bibinfo{journal}{J. Chem. Phys.} \textbf{\bibinfo{volume}{133}},
  \bibinfo{pages}{241103} (\bibinfo{year}{2010}).

\bibitem[{\citenamefont{{Ralph G. Pearson}}(1987)}]{MaxHardnessPearson}
\bibinfo{author}{\bibnamefont{{Ralph G. Pearson}}}, \bibinfo{journal}{J.
  Climate} \textbf{\bibinfo{volume}{64}}, \bibinfo{pages}{561}
  (\bibinfo{year}{1987}).

\bibitem[{\citenamefont{Mezey}(1985)}]{ConcavityMezey1985}
\bibinfo{author}{\bibfnamefont{P.~G.} \bibnamefont{Mezey}},
  \bibinfo{journal}{J. Am. Chem. Soc.} \textbf{\bibinfo{volume}{107}},
  \bibinfo{pages}{3100} (\bibinfo{year}{1985}).

\bibitem[{\citenamefont{Feynman}(1939)}]{HF}
\bibinfo{author}{\bibfnamefont{R.~P.} \bibnamefont{Feynman}},
  \bibinfo{journal}{Phys. Rev.} \textbf{\bibinfo{volume}{56}},
  \bibinfo{pages}{340} (\bibinfo{year}{1939}).

\bibitem[{\citenamefont{von Lilienfeld}(2009)}]{anatole-jcp2009-2}
\bibinfo{author}{\bibfnamefont{O.~A.} \bibnamefont{von Lilienfeld}},
  \bibinfo{journal}{J. Chem. Phys.} \textbf{\bibinfo{volume}{131}},
  \bibinfo{pages}{164102} (\bibinfo{year}{2009}).

\bibitem[{\citenamefont{Smith and van
  Gunsteren}(1994)}]{HigherOrderAlchemicalDerivatives_SmithGunsteren1994}
\bibinfo{author}{\bibfnamefont{P.~E.} \bibnamefont{Smith}} \bibnamefont{and}
  \bibinfo{author}{\bibfnamefont{W.~F.} \bibnamefont{van Gunsteren}},
  \bibinfo{journal}{J. Chem. Phys.} \textbf{\bibinfo{volume}{100}},
  \bibinfo{pages}{577} (\bibinfo{year}{1994}).

\bibitem[{\citenamefont{Putrino et~al.}(2000)\citenamefont{Putrino, Sebastiani,
  and Parrinello}}]{apdsmp}
\bibinfo{author}{\bibfnamefont{A.}~\bibnamefont{Putrino}},
  \bibinfo{author}{\bibfnamefont{D.}~\bibnamefont{Sebastiani}},
  \bibnamefont{and}
  \bibinfo{author}{\bibfnamefont{M.}~\bibnamefont{Parrinello}},
  \bibinfo{journal}{J. Chem. Phys.} \textbf{\bibinfo{volume}{113}},
  \bibinfo{pages}{7102} (\bibinfo{year}{2000}).

\bibitem[{\citenamefont{Beste et~al.}(2006)\citenamefont{Beste, Harrison, and
  Yanai}}]{ArianaAlchemy2006}
\bibinfo{author}{\bibfnamefont{A.}~\bibnamefont{Beste}},
  \bibinfo{author}{\bibfnamefont{R.~J.} \bibnamefont{Harrison}},
  \bibnamefont{and} \bibinfo{author}{\bibfnamefont{T.}~\bibnamefont{Yanai}},
  \bibinfo{journal}{J. Phys. Chem.} \textbf{\bibinfo{volume}{125}},
  \bibinfo{pages}{074101} (\bibinfo{year}{2006}).

\bibitem[{\citenamefont{P\'erez and von Lilienfeld}(2011)}]{alejandro-jctc2011}
\bibinfo{author}{\bibfnamefont{A.}~\bibnamefont{P\'erez}} \bibnamefont{and}
  \bibinfo{author}{\bibfnamefont{O.~A.} \bibnamefont{von Lilienfeld}},
  \bibinfo{journal}{J. Chem. Theory Comput.} \textbf{\bibinfo{volume}{7}},
  \bibinfo{pages}{2358} (\bibinfo{year}{2011}).

\bibitem[{\citenamefont{Lin et~al.}(2007{\natexlab{a}})\citenamefont{Lin,
  Coutinho-Neto, Felsenheimer, von Lilienfeld, Tavernelli, and
  Rothlisberger}}]{anatole-prb2007}
\bibinfo{author}{\bibfnamefont{I.-C.} \bibnamefont{Lin}},
  \bibinfo{author}{\bibfnamefont{M.~D.} \bibnamefont{Coutinho-Neto}},
  \bibinfo{author}{\bibfnamefont{C.}~\bibnamefont{Felsenheimer}},
  \bibinfo{author}{\bibfnamefont{O.~A.} \bibnamefont{von Lilienfeld}},
  \bibinfo{author}{\bibfnamefont{I.}~\bibnamefont{Tavernelli}},
  \bibnamefont{and}
  \bibinfo{author}{\bibfnamefont{U.}~\bibnamefont{Rothlisberger}},
  \bibinfo{journal}{Phys. Rev. B} \textbf{\bibinfo{volume}{75}},
  \bibinfo{pages}{205131} (\bibinfo{year}{2007}{\natexlab{a}}).

\bibitem[{\citenamefont{Stiborov\'a et~al.}(2004)\citenamefont{Stiborov\'a,
  Sejbal, Borek-Dohalska, Aimova, Poljakova, Forsterova, Rupertova, Wiesner,
  Hudecek, Wiessler et~al.}}]{EllipticineMutants_Frei2004}
\bibinfo{author}{\bibfnamefont{M.}~\bibnamefont{Stiborov\'a}},
  \bibinfo{author}{\bibfnamefont{J.}~\bibnamefont{Sejbal}},
  \bibinfo{author}{\bibfnamefont{L.}~\bibnamefont{Borek-Dohalska}},
  \bibinfo{author}{\bibfnamefont{D.}~\bibnamefont{Aimova}},
  \bibinfo{author}{\bibfnamefont{J.}~\bibnamefont{Poljakova}},
  \bibinfo{author}{\bibfnamefont{K.}~\bibnamefont{Forsterova}},
  \bibinfo{author}{\bibfnamefont{M.}~\bibnamefont{Rupertova}},
  \bibinfo{author}{\bibfnamefont{J.}~\bibnamefont{Wiesner}},
  \bibinfo{author}{\bibfnamefont{J.}~\bibnamefont{Hudecek}},
  \bibinfo{author}{\bibfnamefont{M.}~\bibnamefont{Wiessler}},
  \bibnamefont{et~al.}, \bibinfo{journal}{Cancer Res.}
  \textbf{\bibinfo{volume}{64}}, \bibinfo{pages}{8374} (\bibinfo{year}{2004}).

\bibitem[{\citenamefont{Mousset et~al.}(2010)\citenamefont{Mousset, Rabot,
  Bouyssou, Coudert, and Gillaizeau}}]{EllipticineBenzoxazinicAnalogues}
\bibinfo{author}{\bibfnamefont{D.}~\bibnamefont{Mousset}},
  \bibinfo{author}{\bibfnamefont{R.}~\bibnamefont{Rabot}},
  \bibinfo{author}{\bibfnamefont{P.}~\bibnamefont{Bouyssou}},
  \bibinfo{author}{\bibfnamefont{G.}~\bibnamefont{Coudert}}, \bibnamefont{and}
  \bibinfo{author}{\bibfnamefont{I.}~\bibnamefont{Gillaizeau}},
  \bibinfo{journal}{Tetrahedron Lett.} \textbf{\bibinfo{volume}{51}},
  \bibinfo{pages}{3987} (\bibinfo{year}{2010}).

\bibitem[{\citenamefont{Canals et~al.}(2005)\citenamefont{Canals, Purciolas,
  Aymami, and Coll}}]{EllipticineXray_2005}
\bibinfo{author}{\bibfnamefont{A.}~\bibnamefont{Canals}},
  \bibinfo{author}{\bibfnamefont{M.}~\bibnamefont{Purciolas}},
  \bibinfo{author}{\bibfnamefont{J.}~\bibnamefont{Aymami}}, \bibnamefont{and}
  \bibinfo{author}{\bibfnamefont{M.}~\bibnamefont{Coll}},
  \bibinfo{journal}{Acta Cryst.} \textbf{\bibinfo{volume}{D61}},
  \bibinfo{pages}{1009} (\bibinfo{year}{2005}).

\bibitem[{\citenamefont{Bissantz et~al.}(2010)\citenamefont{Bissantz, Kuhn, and
  Stahl}}]{BiomolecularInteractions_Stahl2010}
\bibinfo{author}{\bibfnamefont{C.}~\bibnamefont{Bissantz}},
  \bibinfo{author}{\bibfnamefont{B.}~\bibnamefont{Kuhn}}, \bibnamefont{and}
  \bibinfo{author}{\bibfnamefont{M.}~\bibnamefont{Stahl}}, \bibinfo{journal}{J.
  Medic. Chem.} \textbf{\bibinfo{volume}{53}}, \bibinfo{pages}{5061}
  (\bibinfo{year}{2010}).

\bibitem[{\citenamefont{Kolar et~al.}(2010)\citenamefont{Kolar, Kubar, and
  Hobza}}]{EllipticineEntropy_Hobza2010}
\bibinfo{author}{\bibfnamefont{M.}~\bibnamefont{Kolar}},
  \bibinfo{author}{\bibfnamefont{T.}~\bibnamefont{Kubar}}, \bibnamefont{and}
  \bibinfo{author}{\bibfnamefont{P.}~\bibnamefont{Hobza}}, \bibinfo{journal}{J.
  Phys. Chem. B} \textbf{\bibinfo{volume}{114}}, \bibinfo{pages}{13446}
  (\bibinfo{year}{2010}).

\bibitem[{\citenamefont{Tidor}(1993)}]{Lambdaspacetidor}
\bibinfo{author}{\bibfnamefont{B.}~\bibnamefont{Tidor}}, \bibinfo{journal}{J.
  Phys. Chem.} \textbf{\bibinfo{volume}{97}}, \bibinfo{pages}{1069}
  (\bibinfo{year}{1993}).

\bibitem[{\citenamefont{Oostenbrink and van
  Gunsteren}(2005)}]{OostenbrinkReferencCMPD2005}
\bibinfo{author}{\bibfnamefont{C.}~\bibnamefont{Oostenbrink}} \bibnamefont{and}
  \bibinfo{author}{\bibfnamefont{W.~F.} \bibnamefont{van Gunsteren}},
  \bibinfo{journal}{Proc. Natl. Acad. Sci. USA} \textbf{\bibinfo{volume}{102}},
  \bibinfo{pages}{6750} (\bibinfo{year}{2005}).

\bibitem[{\citenamefont{Oostenbrink}(2009)}]{OostenbrinkHostGuest2008}
\bibinfo{author}{\bibfnamefont{C.}~\bibnamefont{Oostenbrink}},
  \bibinfo{journal}{J. Comp. Chem.} \textbf{\bibinfo{volume}{30}},
  \bibinfo{pages}{212} (\bibinfo{year}{2009}).

\bibitem[{\citenamefont{Iftimie et~al.}(2005)\citenamefont{Iftimie, Minary, and
  Tuckerman}}]{AIMD_PNAS_TUCKERMAN2005}
\bibinfo{author}{\bibfnamefont{R.}~\bibnamefont{Iftimie}},
  \bibinfo{author}{\bibfnamefont{P.}~\bibnamefont{Minary}}, \bibnamefont{and}
  \bibinfo{author}{\bibfnamefont{M.~E.} \bibnamefont{Tuckerman}},
  \bibinfo{journal}{Proc. Natl. Acad. Sci. USA} \textbf{\bibinfo{volume}{102}},
  \bibinfo{pages}{6654} (\bibinfo{year}{2005}).

\bibitem[{\citenamefont{Laio et~al.}(2002)\citenamefont{Laio, VandeVondele, and
  Rothlisberger}}]{laio-qmmm}
\bibinfo{author}{\bibfnamefont{A.}~\bibnamefont{Laio}},
  \bibinfo{author}{\bibfnamefont{J.}~\bibnamefont{VandeVondele}},
  \bibnamefont{and}
  \bibinfo{author}{\bibfnamefont{U.}~\bibnamefont{Rothlisberger}},
  \bibinfo{journal}{J. Chem. Phys.} \textbf{\bibinfo{volume}{116}},
  \bibinfo{pages}{6941} (\bibinfo{year}{2002}).

\bibitem[{\citenamefont{Lin et~al.}(2007{\natexlab{b}})\citenamefont{Lin, von
  Lilienfeld, Coutinho-Neto, Tavernelli, and Rothlisberger}}]{anatole-jpcb2007}
\bibinfo{author}{\bibfnamefont{I.-C.} \bibnamefont{Lin}},
  \bibinfo{author}{\bibfnamefont{O.~A.} \bibnamefont{von Lilienfeld}},
  \bibinfo{author}{\bibfnamefont{M.~D.} \bibnamefont{Coutinho-Neto}},
  \bibinfo{author}{\bibfnamefont{I.}~\bibnamefont{Tavernelli}},
  \bibnamefont{and}
  \bibinfo{author}{\bibfnamefont{U.}~\bibnamefont{Rothlisberger}},
  \bibinfo{journal}{J. Phys. Chem. B} \textbf{\bibinfo{volume}{111}},
  \bibinfo{pages}{14346} (\bibinfo{year}{2007}{\natexlab{b}}).

\bibitem[{\citenamefont{von Lilienfeld and Tkatchenko}(2010)}]{anatole-jcp2010}
\bibinfo{author}{\bibfnamefont{O.~A.} \bibnamefont{von Lilienfeld}}
  \bibnamefont{and}
  \bibinfo{author}{\bibfnamefont{A.}~\bibnamefont{Tkatchenko}},
  \bibinfo{journal}{J. Chem. Phys.} \textbf{\bibinfo{volume}{132}},
  \bibinfo{pages}{234109} (\bibinfo{year}{2010}).

\bibitem[{\citenamefont{DiStasio et~al.}(2012)\citenamefont{DiStasio, von
  Lilienfeld, and Tkatchenko}}]{mbd_PNAS2012}
\bibinfo{author}{\bibfnamefont{R.~A.} \bibnamefont{DiStasio}},
  \bibinfo{author}{\bibfnamefont{O.~A.} \bibnamefont{von Lilienfeld}},
  \bibnamefont{and}
  \bibinfo{author}{\bibfnamefont{A.}~\bibnamefont{Tkatchenko}},
  \bibinfo{journal}{Proc. Natl. Acad. Sci. USA} \textbf{\bibinfo{volume}{109}},
  \bibinfo{pages}{14791} (\bibinfo{year}{2012}).

\bibitem[{\citenamefont{Prize}(2011)}]{vonLilienfeldPrize}
\bibinfo{author}{\bibnamefont{Prize}} (\bibinfo{year}{2011}), \bibinfo{note}{{A
  prize award of the equivalent of an ounce of gold was announced during the
  Navigating Chemical Compound Space program in spring 2011 at the Institute of
  Pure and Applied Mathematics, UCLA. The prize is for finding an interpolating
  transform of two iso-electronic {H}amiltonians such that the potential energy
  becomes linear in the interpolating order parameter. The ounce of gold is
  currently held in the form of 100 shares of iShares Trust fund
  (NYSEARCA:IAU), and will be dispensed in cash at instantaneous exchange rate
  upon recognition of a valid solution by a prize-committee. Apart from the
  author, the prize-committee consists of Profs.~K.~Burke, G.~Henkelman,
  K.~R.~M\"uller, and M.~E.~Tuckerman. Contact the author regarding donations
  to increase the prize award. For more information, see
  \texttt{http://www.alcf.anl.gov/$\sim$anatole}.}}

\bibitem[{\citenamefont{Hammett}(1935)}]{Hammett_cr1935}
\bibinfo{author}{\bibfnamefont{L.~P.} \bibnamefont{Hammett}},
  \bibinfo{journal}{Chem. Rev.} \textbf{\bibinfo{volume}{17}},
  \bibinfo{pages}{125} (\bibinfo{year}{1935}).

\bibitem[{\citenamefont{Hammett}(1937)}]{Hammett_jacs1937}
\bibinfo{author}{\bibfnamefont{L.~P.} \bibnamefont{Hammett}},
  \bibinfo{journal}{J. Am. Chem. Soc.} \textbf{\bibinfo{volume}{59}},
  \bibinfo{pages}{96} (\bibinfo{year}{1937}).

\bibitem[{\citenamefont{Pettifor}(1986)}]{StructureMaps_Pettifor1986}
\bibinfo{author}{\bibfnamefont{D.~G.} \bibnamefont{Pettifor}},
  \bibinfo{journal}{J. Phys. C: Solid State Phys.}
  \textbf{\bibinfo{volume}{19}}, \bibinfo{pages}{285} (\bibinfo{year}{1986}).

\bibitem[{\citenamefont{Gonthier et~al.}(2012)\citenamefont{Gonthier,
  Steinmann, Woodrich, and
  Corminboeuf}}]{FuzzyChemicalConcepts_Corminboeuf2012}
\bibinfo{author}{\bibfnamefont{J.~F.} \bibnamefont{Gonthier}},
  \bibinfo{author}{\bibfnamefont{S.~N.} \bibnamefont{Steinmann}},
  \bibinfo{author}{\bibfnamefont{M.~D.} \bibnamefont{Woodrich}},
  \bibnamefont{and}
  \bibinfo{author}{\bibfnamefont{C.}~\bibnamefont{Corminboeuf}},
  \bibinfo{journal}{Chem. Soc. Rev.} \textbf{\bibinfo{volume}{41}},
  \bibinfo{pages}{4671} (\bibinfo{year}{2012}).

\bibitem[{\citenamefont{Hastie et~al.}(2001)\citenamefont{Hastie, Tibshirani,
  and Friedman}}]{HasTibFri01}
\bibinfo{author}{\bibfnamefont{T.}~\bibnamefont{Hastie}},
  \bibinfo{author}{\bibfnamefont{R.}~\bibnamefont{Tibshirani}},
  \bibnamefont{and} \bibinfo{author}{\bibfnamefont{J.}~\bibnamefont{Friedman}},
  \emph{\bibinfo{title}{The Elements of Statistical Learning: data mining,
  inference and prediction}}, Springer series in statistics
  (\bibinfo{publisher}{Springer}, \bibinfo{address}{New York, N.Y.},
  \bibinfo{year}{2001}).

\bibitem[{\citenamefont{Sch{\"o}lkopf and Smola}(2002)}]{MLbook}
\bibinfo{author}{\bibfnamefont{B.}~\bibnamefont{Sch{\"o}lkopf}}
  \bibnamefont{and} \bibinfo{author}{\bibfnamefont{A.~J.} \bibnamefont{Smola}},
  \emph{\bibinfo{title}{Learning with Kernels}} (\bibinfo{publisher}{MIT
  Press}, \bibinfo{address}{Cambridge}, \bibinfo{year}{2002}).

\bibitem[{\citenamefont{Hastie et~al.}(2009)\citenamefont{Hastie, Tibshirani,
  and Friedman}}]{htf2009}
\bibinfo{author}{\bibfnamefont{T.}~\bibnamefont{Hastie}},
  \bibinfo{author}{\bibfnamefont{R.}~\bibnamefont{Tibshirani}},
  \bibnamefont{and} \bibinfo{author}{\bibfnamefont{J.}~\bibnamefont{Friedman}},
  \emph{\bibinfo{title}{The Elements of Statistical Learning. Data Mining,
  Inference, and Prediction}} (\bibinfo{publisher}{Springer},
  \bibinfo{address}{New York}, \bibinfo{year}{2009}), \bibinfo{edition}{2nd}
  ed.

\bibitem[{\citenamefont{M\"uller et~al.}(2001)\citenamefont{M\"uller, Mika,
  R{\"a}tsch, Tsuda, and Sch{\"o}lkopf}}]{MueMikRaeTsuSch01}
\bibinfo{author}{\bibfnamefont{K.-R.} \bibnamefont{M\"uller}},
  \bibinfo{author}{\bibfnamefont{S.}~\bibnamefont{Mika}},
  \bibinfo{author}{\bibfnamefont{G.}~\bibnamefont{R{\"a}tsch}},
  \bibinfo{author}{\bibfnamefont{K.}~\bibnamefont{Tsuda}}, \bibnamefont{and}
  \bibinfo{author}{\bibfnamefont{B.}~\bibnamefont{Sch{\"o}lkopf}},
  \bibinfo{journal}{IEEE Transactions on Neural Networks}
  \textbf{\bibinfo{volume}{12}}, \bibinfo{pages}{181} (\bibinfo{year}{2001}).

\bibitem[{\citenamefont{Cs{\'a}nyi et~al.}(2004)\citenamefont{Cs{\'a}nyi,
  Albaret, Payne, and Vita}}]{lotf2004}
\bibinfo{author}{\bibfnamefont{G.}~\bibnamefont{Cs{\'a}nyi}},
  \bibinfo{author}{\bibfnamefont{T.}~\bibnamefont{Albaret}},
  \bibinfo{author}{\bibfnamefont{M.~C.} \bibnamefont{Payne}}, \bibnamefont{and}
  \bibinfo{author}{\bibfnamefont{A.~D.} \bibnamefont{Vita}},
  \bibinfo{journal}{Phys. Rev. Lett.} \textbf{\bibinfo{volume}{93}},
  \bibinfo{pages}{175503} (\bibinfo{year}{2004}).

\bibitem[{\citenamefont{Maurer et~al.}(2007)\citenamefont{Maurer, Laio,
  Hugosson, Colombo, and Rothlisberger}}]{PatrickForceMatchingJCTC2007}
\bibinfo{author}{\bibfnamefont{P.}~\bibnamefont{Maurer}},
  \bibinfo{author}{\bibfnamefont{A.}~\bibnamefont{Laio}},
  \bibinfo{author}{\bibfnamefont{H.~W.} \bibnamefont{Hugosson}},
  \bibinfo{author}{\bibfnamefont{M.~C.} \bibnamefont{Colombo}},
  \bibnamefont{and}
  \bibinfo{author}{\bibfnamefont{U.}~\bibnamefont{Rothlisberger}},
  \bibinfo{journal}{J. Chem. Theory Comput.} \textbf{\bibinfo{volume}{3}},
  \bibinfo{pages}{628} (\bibinfo{year}{2007}).

\bibitem[{\citenamefont{Wang and Voorhis}(2010)}]{VoorhisForceMatchingJCTC2010}
\bibinfo{author}{\bibfnamefont{L.~P.} \bibnamefont{Wang}} \bibnamefont{and}
  \bibinfo{author}{\bibfnamefont{T.~V.} \bibnamefont{Voorhis}},
  \bibinfo{journal}{J. Chem. Phys.} \textbf{\bibinfo{volume}{133}},
  \bibinfo{pages}{231101} (\bibinfo{year}{2010}).

\bibitem[{\citenamefont{Hu et~al.}(2003)\citenamefont{Hu, Wang, Wong, and
  Chen}}]{NN4B3LYP_Chen2003}
\bibinfo{author}{\bibfnamefont{L.}~\bibnamefont{Hu}},
  \bibinfo{author}{\bibfnamefont{X.}~\bibnamefont{Wang}},
  \bibinfo{author}{\bibfnamefont{L.}~\bibnamefont{Wong}}, \bibnamefont{and}
  \bibinfo{author}{\bibfnamefont{G.}~\bibnamefont{Chen}}, \bibinfo{journal}{J.
  Chem. Phys.} \textbf{\bibinfo{volume}{119}}, \bibinfo{pages}{11501}
  (\bibinfo{year}{2003}).

\bibitem[{\citenamefont{Zheng et~al.}(2004)\citenamefont{Zheng, Hu, Wang, and
  Chen}}]{NN4B3LYP_Chen2004}
\bibinfo{author}{\bibfnamefont{X.}~\bibnamefont{Zheng}},
  \bibinfo{author}{\bibfnamefont{L.}~\bibnamefont{Hu}},
  \bibinfo{author}{\bibfnamefont{X.}~\bibnamefont{Wang}}, \bibnamefont{and}
  \bibinfo{author}{\bibfnamefont{G.}~\bibnamefont{Chen}},
  \bibinfo{journal}{Chem. Phys. Lett.} \textbf{\bibinfo{volume}{390}},
  \bibinfo{pages}{186} (\bibinfo{year}{2004}).

\bibitem[{\citenamefont{Brown et~al.}(2003)\citenamefont{Brown, Braams,
  Christoffel, Jin, and Bowman}}]{PotentialEnergyFit_Bowman2003}
\bibinfo{author}{\bibfnamefont{A.}~\bibnamefont{Brown}},
  \bibinfo{author}{\bibfnamefont{B.~J.} \bibnamefont{Braams}},
  \bibinfo{author}{\bibfnamefont{K.}~\bibnamefont{Christoffel}},
  \bibinfo{author}{\bibfnamefont{Z.}~\bibnamefont{Jin}}, \bibnamefont{and}
  \bibinfo{author}{\bibfnamefont{J.~M.} \bibnamefont{Bowman}},
  \bibinfo{journal}{J. Chem. Phys.} \textbf{\bibinfo{volume}{119}},
  \bibinfo{pages}{8790} (\bibinfo{year}{2003}).

\bibitem[{\citenamefont{Lorenz et~al.}(2004)\citenamefont{Lorenz, Gross, and
  Scheffler}}]{Neuralnetworks_Scheffler2004}
\bibinfo{author}{\bibfnamefont{S.}~\bibnamefont{Lorenz}},
  \bibinfo{author}{\bibfnamefont{A.}~\bibnamefont{Gross}}, \bibnamefont{and}
  \bibinfo{author}{\bibfnamefont{M.}~\bibnamefont{Scheffler}},
  \bibinfo{journal}{Chem. Phys. Lett.} \textbf{\bibinfo{volume}{395}},
  \bibinfo{pages}{210} (\bibinfo{year}{2004}).

\bibitem[{\citenamefont{Behler and
  Parrinello}(2007)}]{Neuralnetworks_BehlerParrinello2007}
\bibinfo{author}{\bibfnamefont{J.}~\bibnamefont{Behler}} \bibnamefont{and}
  \bibinfo{author}{\bibfnamefont{M.}~\bibnamefont{Parrinello}},
  \bibinfo{journal}{Phys. Rev. Lett.} \textbf{\bibinfo{volume}{98}},
  \bibinfo{pages}{146401} (\bibinfo{year}{2007}).

\bibitem[{\citenamefont{Handley and
  Popelier}(2009)}]{MachineLearningWaterPotential_Handley2008}
\bibinfo{author}{\bibfnamefont{C.~M.} \bibnamefont{Handley}} \bibnamefont{and}
  \bibinfo{author}{\bibfnamefont{P.~L.~A.} \bibnamefont{Popelier}},
  \bibinfo{journal}{J. Chem. Theory Comput.} \textbf{\bibinfo{volume}{5}},
  \bibinfo{pages}{1474} (\bibinfo{year}{2009}).

\bibitem[{\citenamefont{Behler et~al.}(2008)\citenamefont{Behler, Martonak,
  Donadio, and Parrinello}}]{Neuralnetworks_Behler2008}
\bibinfo{author}{\bibfnamefont{J.}~\bibnamefont{Behler}},
  \bibinfo{author}{\bibfnamefont{R.}~\bibnamefont{Martonak}},
  \bibinfo{author}{\bibfnamefont{D.}~\bibnamefont{Donadio}}, \bibnamefont{and}
  \bibinfo{author}{\bibfnamefont{M.}~\bibnamefont{Parrinello}},
  \bibinfo{journal}{Phys. Rev. Lett.} \textbf{\bibinfo{volume}{100}},
  \bibinfo{pages}{185501} (\bibinfo{year}{2008}).

\bibitem[{\citenamefont{Behler}(2011{\natexlab{a}})}]{b2011d}
\bibinfo{author}{\bibfnamefont{J.}~\bibnamefont{Behler}},
  \bibinfo{journal}{Phys. Chem. Chem. Phys.} \textbf{\bibinfo{volume}{13}},
  \bibinfo{pages}{17930} (\bibinfo{year}{2011}{\natexlab{a}}).

\bibitem[{\citenamefont{Balabin and Lomakina}(2011)}]{SVM4CBS_Lomakina2011}
\bibinfo{author}{\bibfnamefont{R.~M.} \bibnamefont{Balabin}} \bibnamefont{and}
  \bibinfo{author}{\bibfnamefont{E.~I.} \bibnamefont{Lomakina}},
  \bibinfo{journal}{Phys. Chem. Chem. Phys.} \textbf{\bibinfo{volume}{13}},
  \bibinfo{pages}{11710} (\bibinfo{year}{2011}).

\bibitem[{\citenamefont{Bart{\'o}k et~al.}(2010)\citenamefont{Bart{\'o}k,
  Payne, Kondor, and Cs{\'a}nyi}}]{bpkc2010}
\bibinfo{author}{\bibfnamefont{A.~P.} \bibnamefont{Bart{\'o}k}},
  \bibinfo{author}{\bibfnamefont{M.~C.} \bibnamefont{Payne}},
  \bibinfo{author}{\bibfnamefont{R.}~\bibnamefont{Kondor}}, \bibnamefont{and}
  \bibinfo{author}{\bibfnamefont{G.}~\bibnamefont{Cs{\'a}nyi}},
  \bibinfo{journal}{Phys. Rev. Lett.} \textbf{\bibinfo{volume}{104}},
  \bibinfo{pages}{136403} (\bibinfo{year}{2010}).

\bibitem[{\citenamefont{Curtarolo et~al.}(2003)\citenamefont{Curtarolo, Morgan,
  Persson, Rodgers, and Ceder}}]{CurtaroloPRL2003}
\bibinfo{author}{\bibfnamefont{S.}~\bibnamefont{Curtarolo}},
  \bibinfo{author}{\bibfnamefont{D.}~\bibnamefont{Morgan}},
  \bibinfo{author}{\bibfnamefont{K.}~\bibnamefont{Persson}},
  \bibinfo{author}{\bibfnamefont{J.}~\bibnamefont{Rodgers}}, \bibnamefont{and}
  \bibinfo{author}{\bibfnamefont{G.}~\bibnamefont{Ceder}},
  \bibinfo{journal}{Phys. Rev. Lett.} \textbf{\bibinfo{volume}{91}},
  \bibinfo{pages}{135503} (\bibinfo{year}{2003}).

\bibitem[{\citenamefont{Setyawan and Curtarolo}(2010)}]{CurtaroloCMS2010}
\bibinfo{author}{\bibfnamefont{W.}~\bibnamefont{Setyawan}} \bibnamefont{and}
  \bibinfo{author}{\bibfnamefont{S.}~\bibnamefont{Curtarolo}},
  \bibinfo{journal}{Comp. Mat. Sci.} \textbf{\bibinfo{volume}{49}},
  \bibinfo{pages}{299} (\bibinfo{year}{2010}).

\bibitem[{\citenamefont{Hautier et~al.}(2010)\citenamefont{Hautier, Fischer,
  Jain, Mueller, and Ceder}}]{MachineLearningHautierCeder2010}
\bibinfo{author}{\bibfnamefont{G.}~\bibnamefont{Hautier}},
  \bibinfo{author}{\bibfnamefont{C.~C.} \bibnamefont{Fischer}},
  \bibinfo{author}{\bibfnamefont{A.}~\bibnamefont{Jain}},
  \bibinfo{author}{\bibfnamefont{T.}~\bibnamefont{Mueller}}, \bibnamefont{and}
  \bibinfo{author}{\bibfnamefont{G.}~\bibnamefont{Ceder}},
  \bibinfo{journal}{Chem. Mater.} \textbf{\bibinfo{volume}{22}},
  \bibinfo{pages}{3762} (\bibinfo{year}{2010}).

\bibitem[{\citenamefont{Hutchison et~al.}(2005)\citenamefont{Hutchison, Ratner,
  and Marks}}]{RatnerJACS2005}
\bibinfo{author}{\bibfnamefont{G.~R.} \bibnamefont{Hutchison}},
  \bibinfo{author}{\bibfnamefont{M.~A.} \bibnamefont{Ratner}},
  \bibnamefont{and} \bibinfo{author}{\bibfnamefont{T.~J.} \bibnamefont{Marks}},
  \bibinfo{journal}{J. Am. Chem. Soc.} \textbf{\bibinfo{volume}{127}},
  \bibinfo{pages}{2339} (\bibinfo{year}{2005}).

\bibitem[{\citenamefont{Misra et~al.}(2011)\citenamefont{Misra, Andrienko,
  Baumeier, Faulon, and von Lilienfeld}}]{anatole-MilindDenis2011}
\bibinfo{author}{\bibfnamefont{M.}~\bibnamefont{Misra}},
  \bibinfo{author}{\bibfnamefont{D.}~\bibnamefont{Andrienko}},
  \bibinfo{author}{\bibfnamefont{B.}~\bibnamefont{Baumeier}},
  \bibinfo{author}{\bibfnamefont{J.-L.} \bibnamefont{Faulon}},
  \bibnamefont{and} \bibinfo{author}{\bibfnamefont{O.~A.} \bibnamefont{von
  Lilienfeld}}, \bibinfo{journal}{J. Chem. Theory Comput.}
  \textbf{\bibinfo{volume}{7}}, \bibinfo{pages}{2549} (\bibinfo{year}{2011}).

\bibitem[{\citenamefont{Snyder et~al.}(2012)\citenamefont{Snyder, Rupp, Hansen,
  M\"uller, and Burke}}]{ML4Kieron2012}
\bibinfo{author}{\bibfnamefont{J.~C.} \bibnamefont{Snyder}},
  \bibinfo{author}{\bibfnamefont{M.}~\bibnamefont{Rupp}},
  \bibinfo{author}{\bibfnamefont{K.}~\bibnamefont{Hansen}},
  \bibinfo{author}{\bibfnamefont{K.-R.} \bibnamefont{M\"uller}},
  \bibnamefont{and} \bibinfo{author}{\bibfnamefont{K.}~\bibnamefont{Burke}},
  \bibinfo{journal}{Phys. Rev. Lett.} \textbf{\bibinfo{volume}{108}},
  \bibinfo{pages}{253002} (\bibinfo{year}{2012}).

\bibitem[{\citenamefont{Pozun et~al.}(2012)\citenamefont{Pozun, Hansen,
  Sheppard, Rupp, M\"uller, and Henkelman}}]{ML4Graeme2012}
\bibinfo{author}{\bibfnamefont{Z.~D.} \bibnamefont{Pozun}},
  \bibinfo{author}{\bibfnamefont{K.}~\bibnamefont{Hansen}},
  \bibinfo{author}{\bibfnamefont{D.}~\bibnamefont{Sheppard}},
  \bibinfo{author}{\bibfnamefont{M.}~\bibnamefont{Rupp}},
  \bibinfo{author}{\bibfnamefont{K.-R.} \bibnamefont{M\"uller}},
  \bibnamefont{and}
  \bibinfo{author}{\bibfnamefont{G.}~\bibnamefont{Henkelman}},
  \bibinfo{journal}{J. Chem. Phys.} \textbf{\bibinfo{volume}{136}},
  \bibinfo{pages}{174101} (\bibinfo{year}{2012}).

\bibitem[{\citenamefont{Wellendorff et~al.}(2012)\citenamefont{Wellendorff,
  Lundgaard, M{\o}gelh{\o}j, Petzold, Landis, N{\o}rskov, Bligaard, and
  Jacobsen}}]{Bayesian4DFT_Jacobsen2012}
\bibinfo{author}{\bibfnamefont{J.}~\bibnamefont{Wellendorff}},
  \bibinfo{author}{\bibfnamefont{K.~T.} \bibnamefont{Lundgaard}},
  \bibinfo{author}{\bibfnamefont{A.}~\bibnamefont{M{\o}gelh{\o}j}},
  \bibinfo{author}{\bibfnamefont{V.}~\bibnamefont{Petzold}},
  \bibinfo{author}{\bibfnamefont{D.~D.} \bibnamefont{Landis}},
  \bibinfo{author}{\bibfnamefont{J.~K.} \bibnamefont{N{\o}rskov}},
  \bibinfo{author}{\bibfnamefont{T.}~\bibnamefont{Bligaard}}, \bibnamefont{and}
  \bibinfo{author}{\bibfnamefont{K.~W.} \bibnamefont{Jacobsen}},
  \bibinfo{journal}{Phys. Rev. B} \textbf{\bibinfo{volume}{85}},
  \bibinfo{pages}{235149} (\bibinfo{year}{2012}).

\bibitem[{\citenamefont{Schneider}(2010)}]{SchneiderReview2010}
\bibinfo{author}{\bibfnamefont{G.}~\bibnamefont{Schneider}},
  \bibinfo{journal}{Nature Reviews} \textbf{\bibinfo{volume}{9}},
  \bibinfo{pages}{273} (\bibinfo{year}{2010}).

\bibitem[{\citenamefont{Ivanciuc}(2000)}]{WienerDescriptors}
\bibinfo{author}{\bibfnamefont{O.}~\bibnamefont{Ivanciuc}},
  \bibinfo{journal}{J. Chem. Inf. Comp. Sci.} \textbf{\bibinfo{volume}{40}},
  \bibinfo{pages}{1412} (\bibinfo{year}{2000}).

\bibitem[{\citenamefont{Todeschini and
  Consonni}(2009)}]{TodeschiniConsonniHandbookDescriptor}
\bibinfo{author}{\bibfnamefont{R.}~\bibnamefont{Todeschini}} \bibnamefont{and}
  \bibinfo{author}{\bibfnamefont{V.}~\bibnamefont{Consonni}},
  \emph{\bibinfo{title}{Handbook of Molecular Descriptors}}
  (\bibinfo{publisher}{Wiley-VCH, Weinheim}, \bibinfo{year}{2009}).

\bibitem[{\citenamefont{Braun et~al.}(2005)\citenamefont{Braun, Kerber,
  Meringer, and R\"ucker}}]{DescriptroOverviewMeringer2005}
\bibinfo{author}{\bibfnamefont{J.}~\bibnamefont{Braun}},
  \bibinfo{author}{\bibfnamefont{A.}~\bibnamefont{Kerber}},
  \bibinfo{author}{\bibfnamefont{M.}~\bibnamefont{Meringer}}, \bibnamefont{and}
  \bibinfo{author}{\bibfnamefont{C.}~\bibnamefont{R\"ucker}},
  \bibinfo{journal}{MATCH} \textbf{\bibinfo{volume}{54}}, \bibinfo{pages}{163}
  (\bibinfo{year}{2005}).

\bibitem[{\citenamefont{Faulon et~al.}(2003)\citenamefont{Faulon, {Visco, Jr.},
  and Pophale}}]{SignatureFaulon2003}
\bibinfo{author}{\bibfnamefont{J.-L.} \bibnamefont{Faulon}},
  \bibinfo{author}{\bibfnamefont{D.~P.} \bibnamefont{{Visco, Jr.}}},
  \bibnamefont{and} \bibinfo{author}{\bibfnamefont{R.~S.}
  \bibnamefont{Pophale}}, \bibinfo{journal}{J. Chem. Inf. Comp. Sci.}
  \textbf{\bibinfo{volume}{43}}, \bibinfo{pages}{707} (\bibinfo{year}{2003}).

\bibitem[{\citenamefont{Feng et~al.}(2009)\citenamefont{Feng, Marcon, Pisula,
  Hansen, Kirkpatrick, Grozema, Andrienko, Kremer, and
  Mullen}}]{AndrienkoKremerMullen2009}
\bibinfo{author}{\bibfnamefont{X.}~\bibnamefont{Feng}},
  \bibinfo{author}{\bibfnamefont{V.}~\bibnamefont{Marcon}},
  \bibinfo{author}{\bibfnamefont{W.}~\bibnamefont{Pisula}},
  \bibinfo{author}{\bibfnamefont{M.~R.} \bibnamefont{Hansen}},
  \bibinfo{author}{\bibfnamefont{J.}~\bibnamefont{Kirkpatrick}},
  \bibinfo{author}{\bibfnamefont{F.}~\bibnamefont{Grozema}},
  \bibinfo{author}{\bibfnamefont{D.}~\bibnamefont{Andrienko}},
  \bibinfo{author}{\bibfnamefont{K.}~\bibnamefont{Kremer}}, \bibnamefont{and}
  \bibinfo{author}{\bibfnamefont{K.}~\bibnamefont{Mullen}},
  \bibinfo{journal}{Nature Materials} \textbf{\bibinfo{volume}{8}},
  \bibinfo{pages}{421} (\bibinfo{year}{2009}).

\bibitem[{\citenamefont{Rupp et~al.}(2012{\natexlab{a}})\citenamefont{Rupp,
  Tkatchenko, M\"uller, and von Lilienfeld}}]{RuppPRL2012}
\bibinfo{author}{\bibfnamefont{M.}~\bibnamefont{Rupp}},
  \bibinfo{author}{\bibfnamefont{A.}~\bibnamefont{Tkatchenko}},
  \bibinfo{author}{\bibfnamefont{K.-R.} \bibnamefont{M\"uller}},
  \bibnamefont{and} \bibinfo{author}{\bibfnamefont{O.~A.} \bibnamefont{von
  Lilienfeld}}, \bibinfo{journal}{Phys. Rev. Lett.}
  \textbf{\bibinfo{volume}{108}}, \bibinfo{pages}{058301}
  (\bibinfo{year}{2012}{\natexlab{a}}).

\bibitem[{\citenamefont{Kohn and Sham}(1965)}]{KS}
\bibinfo{author}{\bibfnamefont{W.}~\bibnamefont{Kohn}} \bibnamefont{and}
  \bibinfo{author}{\bibfnamefont{L.~J.} \bibnamefont{Sham}},
  \bibinfo{journal}{Phys. Rev.} \textbf{\bibinfo{volume}{140}},
  \bibinfo{pages}{A1133} (\bibinfo{year}{1965}).

\bibitem[{\citenamefont{Becke}(1993)}]{becke3}
\bibinfo{author}{\bibfnamefont{A.~D.} \bibnamefont{Becke}},
  \bibinfo{journal}{J. Chem. Phys.} \textbf{\bibinfo{volume}{98}},
  \bibinfo{pages}{5648} (\bibinfo{year}{1993}).

\bibitem[{\citenamefont{Perdew et~al.}(1996)\citenamefont{Perdew, Ernzerhof,
  and Burke}}]{PBE0}
\bibinfo{author}{\bibfnamefont{J.~P.} \bibnamefont{Perdew}},
  \bibinfo{author}{\bibfnamefont{M.}~\bibnamefont{Ernzerhof}},
  \bibnamefont{and} \bibinfo{author}{\bibfnamefont{K.}~\bibnamefont{Burke}},
  \bibinfo{journal}{J. Chem. Phys.} \textbf{\bibinfo{volume}{105}},
  \bibinfo{pages}{9982} (\bibinfo{year}{1996}).

\bibitem[{\citenamefont{Ernzerhof and Scuseria}(1999)}]{PBE01}
\bibinfo{author}{\bibfnamefont{M.}~\bibnamefont{Ernzerhof}} \bibnamefont{and}
  \bibinfo{author}{\bibfnamefont{G.~E.} \bibnamefont{Scuseria}},
  \bibinfo{journal}{J. Chem. Phys.} \textbf{\bibinfo{volume}{110}},
  \bibinfo{pages}{5029} (\bibinfo{year}{1999}).

\bibitem[{\citenamefont{Lynch and Truhlar}(2003)}]{DFT4G3_Truhlar2003}
\bibinfo{author}{\bibfnamefont{B.~J.} \bibnamefont{Lynch}} \bibnamefont{and}
  \bibinfo{author}{\bibfnamefont{D.~G.} \bibnamefont{Truhlar}},
  \bibinfo{journal}{J. Phys. Chem. A} \textbf{\bibinfo{volume}{107}},
  \bibinfo{pages}{3898} (\bibinfo{year}{2003}).

\bibitem[{\citenamefont{Montavon et~al.}(2013)\citenamefont{Montavon, Hansen,
  Fazli, Rupp, Biegler, Ziehe, Tkatchenko, von Lilienfeld, and
  M\"uller}}]{Montavon_NIPS2012}
\bibinfo{author}{\bibfnamefont{G.}~\bibnamefont{Montavon}},
  \bibinfo{author}{\bibfnamefont{K.}~\bibnamefont{Hansen}},
  \bibinfo{author}{\bibfnamefont{S.}~\bibnamefont{Fazli}},
  \bibinfo{author}{\bibfnamefont{M.}~\bibnamefont{Rupp}},
  \bibinfo{author}{\bibfnamefont{F.}~\bibnamefont{Biegler}},
  \bibinfo{author}{\bibfnamefont{A.}~\bibnamefont{Ziehe}},
  \bibinfo{author}{\bibfnamefont{A.}~\bibnamefont{Tkatchenko}},
  \bibinfo{author}{\bibfnamefont{O.~A.} \bibnamefont{von Lilienfeld}},
  \bibnamefont{and} \bibinfo{author}{\bibfnamefont{K.-R.}
  \bibnamefont{M\"uller}}, \bibinfo{journal}{NIPS proceedings}
  (\bibinfo{year}{2013}), \bibinfo{note}{accepted}.

\bibitem[{\citenamefont{Montavon et~al.}(2012)\citenamefont{Montavon, Rupp,
  Gobre, Vazquez-Mayagoitia, Hansen, Tkatchenko, M\"uller, and von
  Lilienfeld}}]{Montavon2012}
\bibinfo{author}{\bibfnamefont{G.}~\bibnamefont{Montavon}},
  \bibinfo{author}{\bibfnamefont{M.}~\bibnamefont{Rupp}},
  \bibinfo{author}{\bibfnamefont{V.}~\bibnamefont{Gobre}},
  \bibinfo{author}{\bibfnamefont{A.}~\bibnamefont{Vazquez-Mayagoitia}},
  \bibinfo{author}{\bibfnamefont{K.}~\bibnamefont{Hansen}},
  \bibinfo{author}{\bibfnamefont{A.}~\bibnamefont{Tkatchenko}},
  \bibinfo{author}{\bibfnamefont{K.-R.} \bibnamefont{M\"uller}},
  \bibnamefont{and} \bibinfo{author}{\bibfnamefont{O.~A.} \bibnamefont{von
  Lilienfeld}} (\bibinfo{year}{2012}), \bibinfo{note}{submitted}.

\bibitem[{Foo()}]{FootnoteCoulombMatrix}
\bibinfo{note}{Using 2 as powers of $Z$ (the energy of the hydrogenic atom) did
  not change performance.}

\bibitem[{\citenamefont{Patterson}(1939)}]{NatureHomometric}
\bibinfo{author}{\bibfnamefont{A.~L.} \bibnamefont{Patterson}},
  \bibinfo{journal}{Nature} \textbf{\bibinfo{volume}{143}},
  \bibinfo{pages}{939} (\bibinfo{year}{1939}).

\bibitem[{\citenamefont{Moussa}(2012)}]{MoussaComment}
\bibinfo{author}{\bibfnamefont{J.~E.} \bibnamefont{Moussa}},
  \bibinfo{journal}{Phys. Rev. Lett.} \textbf{\bibinfo{volume}{109}},
  \bibinfo{pages}{059801} (\bibinfo{year}{2012}).

\bibitem[{\citenamefont{Rupp et~al.}(2012{\natexlab{b}})\citenamefont{Rupp,
  Tkatchenko, M\"uller, and von Lilienfeld}}]{MoussaReply}
\bibinfo{author}{\bibfnamefont{M.}~\bibnamefont{Rupp}},
  \bibinfo{author}{\bibfnamefont{A.}~\bibnamefont{Tkatchenko}},
  \bibinfo{author}{\bibfnamefont{K.-R.} \bibnamefont{M\"uller}},
  \bibnamefont{and} \bibinfo{author}{\bibfnamefont{O.~A.} \bibnamefont{von
  Lilienfeld}}, \bibinfo{journal}{Phys. Rev. Lett.}
  \textbf{\bibinfo{volume}{109}}, \bibinfo{pages}{059802}
  (\bibinfo{year}{2012}{\natexlab{b}}).

\bibitem[{\citenamefont{Ciresan et~al.}(2010)\citenamefont{Ciresan, Meier,
  Gambardella, and Schmidhuber}}]{ciresan}
\bibinfo{author}{\bibfnamefont{D.~C.} \bibnamefont{Ciresan}},
  \bibinfo{author}{\bibfnamefont{U.}~\bibnamefont{Meier}},
  \bibinfo{author}{\bibfnamefont{L.~M.} \bibnamefont{Gambardella}},
  \bibnamefont{and}
  \bibinfo{author}{\bibfnamefont{J.}~\bibnamefont{Schmidhuber}},
  \bibinfo{journal}{Neural Computation} \textbf{\bibinfo{volume}{22}},
  \bibinfo{pages}{3207} (\bibinfo{year}{2010}).

\bibitem[{\citenamefont{Behler}(2011{\natexlab{b}})}]{Neuralnetworks_Behler201%
1}
\bibinfo{author}{\bibfnamefont{J.}~\bibnamefont{Behler}}, \bibinfo{journal}{J.
  Chem. Phys.} \textbf{\bibinfo{volume}{134}}, \bibinfo{pages}{074106}
  (\bibinfo{year}{2011}{\natexlab{b}}).

\bibitem[{\citenamefont{Doraiswamy et~al.}(2012)\citenamefont{Doraiswamy,
  Bender, Candler, Paukku, Yang, Varga, and Truhlar}}]{TruhlarHomometric}
\bibinfo{author}{\bibfnamefont{S.}~\bibnamefont{Doraiswamy}},
  \bibinfo{author}{\bibfnamefont{J.}~\bibnamefont{Bender}},
  \bibinfo{author}{\bibfnamefont{G.~V.} \bibnamefont{Candler}},
  \bibinfo{author}{\bibfnamefont{Y.}~\bibnamefont{Paukku}},
  \bibinfo{author}{\bibfnamefont{K.}~\bibnamefont{Yang}},
  \bibinfo{author}{\bibfnamefont{Z.}~\bibnamefont{Varga}}, \bibnamefont{and}
  \bibinfo{author}{\bibfnamefont{D.~G.} \bibnamefont{Truhlar}}
  (\bibinfo{year}{2012}), \bibinfo{note}{to be published}.

\bibitem[{\citenamefont{Chen and Zhou}(2008)}]{Review_orbitalfreeDFT2008}
\bibinfo{author}{\bibfnamefont{H.}~\bibnamefont{Chen}} \bibnamefont{and}
  \bibinfo{author}{\bibfnamefont{A.}~\bibnamefont{Zhou}},
  \bibinfo{journal}{Numer. Math. Theor. Meth. Appl.}
  \textbf{\bibinfo{volume}{1}}, \bibinfo{pages}{1} (\bibinfo{year}{2008}).

\bibitem[{\citenamefont{von Lilienfeld and
  Knoll}(2012)}]{FourierDescriptor_2012}
\bibinfo{author}{\bibfnamefont{O.~A.} \bibnamefont{von Lilienfeld}}
  \bibnamefont{and} \bibinfo{author}{\bibfnamefont{A.}~\bibnamefont{Knoll}}
  (\bibinfo{year}{2012}), \bibinfo{note}{submitted}.

\end{thebibliography}
\end{document}